\documentclass[fleqn,usenatbib]{mnras}

\usepackage{newtxtext,newtxmath}

\usepackage[T1]{fontenc}

\DeclareRobustCommand{\VAN}[3]{#2}
\let\VANthebibliography\thebibliography
\def\thebibliography{\DeclareRobustCommand{\VAN}[3]{##3}\VANthebibliography}

\usepackage{graphicx}	
\usepackage{amsmath}	

\usepackage{graphicx}	
\usepackage{amsmath}	
\usepackage{xcolor}
\usepackage{ulem} 
\usepackage{lipsum}

\newcommand{\eq}[1]{Equation~\ref{eq:#1}}
\newcommand{\fig}[1]{Figure~\ref{fig:#1}}
\renewcommand{\sec}[1]{Section~\ref{sec:#1}}

\newcommand{\Eq}[1]{Equation~\ref{eq:#1}}

\newcommand{\eagle}{\mbox{\sc{Eagle}}}

\newcommand{\flares}{\mbox{\sc Flares}}

\def\mnras{MNRAS}

\def\apj{ApJ}
\def\apjs{ApJS}
\def\apjl{ApJL}

\def\P3M{P$^3$M}

\title[FLARES IX]{FLARES IX: The Physical Mechanisms Driving Compact Galaxy Formation and Evolution}

\author[William J. Roper et al.]{William J. Roper$^{1}$\thanks{E-mail: w.roper@sussex.ac.uk}, 
Christopher C. Lovell$^{2}$, 
Aswin P. Vijayan$^{3,4}$, 
Dimitrios Irodotou$^{5}$, 
\newauthor 
Jussi K. Kuusisto$^{1}$, 
Jasleen Matharu$^{3}$, 
Louise T. C. Seeyave$^{1}$, 
Peter A. Thomas$^{1}$, 
Stephen M. Wilkins$^{1,6}$ 
\newauthor 
\\
$^{1}$Astronomy Centre, University of Sussex, Falmer, Brighton BN1 9QH, UK\\
$^{2}$Institute of Cosmology and Gravitation, University of Portsmouth, Burnaby Road, Portsmouth, PO1 3FX, UK\\
$^{3}$ Cosmic Dawn Center (DAWN) \\
$^{4}$DTU-Space, Technical University of Denmark, Elektrovej 327, DK-2800 Kgs. Lyngby, Denmark \\
$^{5}$Department of Physics, University of Helsinki, Gustaf Hällströmin katu 2, FI-00014, Helsinki, Finland \\
$^{6}$Institute of Space Sciences and Astronomy, University of Malta, Msida MSD 2080, Malta \\
}

\date{Accepted XXX. Received YYY; in original form ZZZ}

\pubyear{2022}

\begin{document}
\label{firstpage}
\pagerange{\pageref{firstpage}--\pageref{lastpage}}
\maketitle

\begin{abstract}
In the \flares\ (First Light And Reionisation Epoch Simulations) suite of hydrodynamical simulations, we find the high redshift ($z>5$) intrinsic size-luminosity relation is, surprisingly, negatively sloped. However, after including the effects of dust attenuation we find a positively sloped UV observed size-luminosity relation in good agreement with other simulated and observational studies. In this work, we extend this analysis to probe the underlying physical mechanisms driving the formation and evolution of the compact galaxies driving the negative size-mass/size-luminosity relation. We find the majority of compact galaxies ($R_{1/2, \star}< 1 \mathrm{pkpc}$), which drive the negative slope of the size-mass relation, have transitioned from extended to compact sizes via efficient centralised cooling, resulting in high specific star formation rates in their cores. These compact stellar systems are enshrouded by non-star forming gas distributions as much as $100\times$ larger than their stellar counterparts. By comparing with galaxies from the \eagle\ simulation suite, we find that these extended gas distributions `turn on' and begin to form stars between $z=5$ and $z=0$ leading to increasing sizes, and thus the evolution of the size-mass relation from a negative to a positive slope. This explicitly demonstrates the process of inside-out galaxy formation in which compact bulges form earlier than the surrounding discs.
\end{abstract}




\section{Introduction}

Galaxy sizes are the macroscopic culmination of a myriad of internal and external multi-scale physical mechanisms, such as galaxy mergers, instabilities, gas accretion, gas transport, star formation and feedback processes \citep{Conselice14}. Many of these processes take place below the resolution of cosmological hydrodynamic simulations that have large enough periodic volumes to yield statistically significant galaxy samples at high redshift. Models describing these small-scale physical mechanisms are so-called `sub-grid' models \citep{SomervilleDave_15}. Galaxy sizes are a powerful diagnostic of the performance of a sub-grid model since, unlike stellar masses or luminosities, they are not only dependent on how much stellar mass is formed but also where this mass is distributed within a galaxy, and thus the properties of the local environment in the Interstellar Medium (ISM). 

The intrinsic size-luminosity relation and the stellar size-mass relation both describe the underlying distribution of stellar mass in galaxies.
At low redshift ($z<2$), simulations produce a positive relation between increasing size and increasing mass. \cite{Furlong_2017} measured the size-mass relation in \eagle\ finding a positive relation which flattens at $z=2$, and an increase in size with decreasing redshift over the range $z=0-2$. Both findings are in good agreement with the observations referenced therein. \cite{Furlong_2017} found that passive galaxies at $z<2$ evolved in size by migration of stellar particles, rather than by star formation or merging mechanisms.

Observations have shown a similar redshift evolution in the size of galaxies from the Epoch of Reionisation to the present day. In the low redshift Universe ($z<3$), star-forming galaxies typically have sizes of the order $1-30$ pkpc \citep{Zhang_2019, Kawinwanichakij_2021}, with a positive size-luminosity relation \citep{van_der_Wel_2014, Suess_2019, Kawinwanichakij_2021}. However, \cite{van_der_Wel_2014} found a substantial population of compact and massive ($R<2$ pkpc, $M_\star>10^{11} \mathrm{M}_\odot$) galaxies at $z=1.5-3$.

At $z>5$ a number of Hubble Space Telescope ({\emph HST}) studies have found bright star-forming galaxies with compact half-light radii of 0.5-1.0 pkpc \citep{Oesch_2010, Grazian_2012, Mosleh_2012, Ono_2013, Huang_2013, Holwerda_2015, Kawamata_2015, Shibuya2015, Kawamata_2018, Holwerda2020}. The aforementioned massive compact galaxies found in \cite{van_der_Wel_2014} could be descendants of these early compact systems that are yet to undergo a process driving size increase, such as stellar migration or inside-out star formation.
These {\emph HST} studies produced size-luminosity relations with positive slopes, but there is appreciable scatter in the reported slopes. The subset of these studies that include lensed sources \citep[e.g.][]{Kawamata_2018} produce steeper slopes and resolve compact galaxies at lower luminosities than accessible to non-lensing surveys, which yield flatter size-luminosity relations.

In \cite{Roper22} we found a negative intrinsic stellar size-luminosity relation at $z>5$ in the First Light And Reionisation Epoch Simulations  \citep[\flares,][]{Lovell2021, Vijayan2020}, a suite of 40 cosmological hydrodynamical zoom simulations using the Evolution and Assembly of GaLaxies and their Environments (\eagle) model \citep{schaye_eagle_2015, crain_eagle_2015}. However, after including the effects of dust attenuation, the observed size-luminosity relation exhibited a positive slope, in good agreement with current {\emph HST} observations \citep{Oesch_2010, Grazian_2012, Mosleh_2012, Ono_2013, Huang_2013, Holwerda_2015, Kawamata_2015, Shibuya2015, Kawamata_2018, Holwerda2020}. This increase in slope and reversal of the trend between size and luminosity was found to be caused by the concentration of compact dust distributions obscuring the brightest regions of compact galaxies. We also probed the size-luminosity relation as a function of wavelength, finding a flattening of the  relation (and indeed a shift to a negative trend) with reddening wavelength, reflecting the underlying size-mass relation. A negative high redshift intrinsic UV size-luminosity/size-mass relation is not unique to the \eagle\ model, with a negative intrinsic stellar size-luminosity relation present in the \textsc{BlueTides} simulation \citep{Feng2016} at $z\geq7$ \citep{Marshall21}, and the \textsc{Illutris-TNG} simulations \citep{Pillepich2018} at $z=5$ \citep{Popping2021}, while a constant or negative stellar size-mass relation, dependent on the measurement method, was found in the \textsc{Simba} simulation \citep{dave_simba:_2019} at $z=6$ \citep{Wu_20}. 

Explicit confirmation or contradiction of a negative high redshift size-mass relation by observations remains an open question. With the infra-red capabilities of JWST we should soon be able to probe the less obscured rest-frame optical emission of galaxies at high redshift. \cite{Roper22} showed it is possible to probe a constant observed size-luminosity relation at $z<5$ using JWST's F444W filter on NIRCam, while MIRI can probe the the negative regime of the observed size-luminosity relation. Using this observational power we will soon ascertain if the negative size-mass relation in simulations is representative of the true Universe. 

The evolution from the compact high redshift Universe to the extended low redshift Universe has been studied extensively in both simulations and observations. For star-forming galaxies at $0.5\lesssim z \lesssim 2.2$, it has been shown that they predominantly grow inside-out via star formation as opposed to growth via merger driven accretion. This trend is particularly evident at the massive end (log($M_{*}\,/\,\mathrm{M_{\odot}}$)$\gtrsim10$), with evidence that they may be concurrently quenching from the inside-out and building bulges \citep{Tacchella_15a, Tacchella_15b, Tacchella_18, Nelson_16, Nelson_19, Wilman_20, Matharu_22}. At $z\sim0$, the picture is less clear, with some evidence of inside-out growth \citep[e.g.][]{Munoz-Mateos_07}, but also a tendency for star-forming regions and stellar disks to be found to have the same size \citep{James_09, Fossati_13}. Upcoming JWST data should be able to complete this picture at least out to $z\sim5$.

Given the surprising high redshift size-mass/luminosity relation results consistently found across various simulations at high redshift, it is imperative we understand the mechanisms that yield these unexpected trends. Regardless of whether future observations confirm or rule out this behaviour, a thorough investigation is necessary to elucidate the processes taking place. Such an investigation will aid the interpretation of observations and form the starting point for modifications to the theoretical models should this behaviour be unique to simulations. In this work, we address this by probing the physical mechanisms driving the formation and evolution of compact galaxies, utilising the environmental coverage of \flares\ to yield a representative description of size evolution from the high redshift Universe ($z>5$) to the present day. 

This article is structured as follows: In \sec{flares} we detail the \flares\ simulations, in \sec{hmr} we discuss stellar and gas half-mass radii and their size-mass relations, in \sec{form} we investigate the formation of compact stellar systems, and in \sec{evolve} we explore the physical mechanisms driving size evolution from the Epoch of Reionisation to the present day. Finally, in \sec{conclusion} we present our conclusions.
Throughout this work, we assume a Planck year 1 cosmology ($\Omega_{m} = 0.307$, $\Omega_\Lambda = 0.693$, $h = 0.6777$, \cite{planck_collaboration_2014}).

\section{First Light And Reionisation Epoch Simulations (\flares)}
\label{sec:flares}

\flares\ is a suite of 40 cosmological hydrodynamical zoom simulations focusing on the Epoch of Reionisation \citep{Lovell2021, Vijayan2020}. Each resimulation is a spherical region with radius 14 cMpc/h, selected from a $(3.2\; \mathrm{cGpc})^{3}$ `parent' dark matter only (DMO) simulation \citep{barnes_redshift_2017}. The size of the parent simulation enables a region selection methodology capturing the rarest most over-dense regions in the Universe where it is thought the most massive galaxies form \citep{chiang_ancient_2013,lovell_characterising_2018}. We select the 40 regions at $z = 4.67$, at which point the most extreme overdensities are only mildly non-linear, thus preserving the overdensity hierarchy across the full resimulation redshift range.
The resimulations span a wide range of overdensities ($\delta=-0.479\to0.970$; see Table A1 of \citealt{Lovell2021}) with a bias towards extreme overdensities to guarantee a statistically significant sample of galaxies accessible to the JWST and other upcoming next generation telescopes such as the Euclid Space Telescope \citep{Euclid_22}, the Nancy Grace Roman Space Telescope \citep{Wang_22}, and the Extremely Large Telescope \citep{ELT_21}.
We store outputs (snapshots) at integer redshifts from $z=15$ to $z=5$, inclusive.

\flares\ employs the AGNdT9 variant of the \eagle\ sub-grid model \citep{schaye_eagle_2015, crain_eagle_2015}. We choose the AGNdT9 variant since it produces similar mass functions to the fiducial model while better reproducing the hot gas of groups and clusters \citep{barnes_cluster-eagle_2017}, environments which are more prevalent in the extreme overdensities focused on in \flares. We adopt the fiducial resolution of the \eagle\ model with a dark matter and an initial gas particle mass of $m_{\mathrm{dm}} = 9.7 \times 10^6\, \mathrm{M}_{\odot}$ and $m_{\mathrm{g}} = 1.8 \times 10^6\, \mathrm{M}_{\odot}$ respectively, and a gravitational softening length of $2.66\, \mathrm{ckpc}$ at $z\geq2.8$. We present a discussion of the features of the \eagle\ model important to this work in \sec{eagle}. 

The \eagle\ model was calibrated using the $z=0$ galaxy mass function, the mass-size relation for discs, and the gas mass-halo mass relation, but is in good agreement with a number of low redshift observables not used for calibration \citep[e.g.][]{furlong_evolution_2015, Trayford2015, Lagos2015}. Despite being calibrated at low redshift the \eagle\ model performs exceptionally well at high redshift. Indeed, previous \flares\ papers have shown a good agreement with: the galaxy stellar mass function \citep{Lovell2021}, the observed UV luminosity function at $z \geq 5$ \citep{Vijayan2020, Vijayan22}, HST constraints on galaxy sizes at $z\geq5$ \citep{Roper22}, and the evolution of galaxy colours with redshift \citep{Wilkins22_color}. In addition, we have also presented galaxy populations at the redshift `frontier' \citep[$z > 10$;][]{Wilkins22_frontier} and the behaviour of star formation and metal enrichment histories during the Epoch of Reionisation \citep{Wilkins22_metal}.

Since the 40 resimulations are both a sub-sample of the environments in the parent DMO simulation and biased to extreme overdensities we apply a statistical weighting scheme to reproduce the true distribution of environments in the universe. We achieve this by applying an overdensity based weighting to each region and then producing composite distribution functions. For a detailed description of this weighting scheme please refer to the introductory \flares\ paper, \cite{Lovell2021}.

\subsection{The \eagle\ Model}
\label{sec:eagle}

The \eagle\ model used in \flares\ is described in detail in \cite{schaye_eagle_2015} and \cite{crain_eagle_2015}; here we give a brief explanation of the elements of the model that are pertinent to this work and refer the reader to the original works for more detail.

\subsubsection{Radiative cooling}
\label{sec:eagle_cooling}

Radiative cooling and photoheating are implemented on an element-by-element basis following \cite{Wiersma2009a} via a look up table. This includes all 11 elements found to be important for radiative cooling and photoheating: H, He, C, N, O, Ne, Mg, Si, S, Ca, and Fe. \cite{Wiersma2009a} tabulated the rates as a function of density, temperature and redshift assuming the gas to be in ionisation equilibrium and exposed to the Cosmic Microwave Background and a UV/X-ray background from galaxies and quasars \citep{haardt_madau2001}. These look up tables were produced using \textsc{cloudy} version 07.02 \citep{Ferland1998}. 

This radiative cooling implementation comes with some caveats that are important to keep in mind in the context of this work. The assumption of ionisation equilibrium and the absence of ionising radiation from local sources may cause overestimated cooling rates in rapidly cooling gas \citep{Oppenheimer13b} and gas that has recently been irradiated by an AGN \cite{Oppenheimer13a}. The model also ignores self-shielding, which could yield underestimated cooling rates in dense gas. Given that much of the star formation during the Epoch of Reionisation happens in pristine dense gas the underestimate due to self-shielding somewhat counteracts the overestimate introduced by assuming ionisation equilibrium in these regions.

\subsubsection{Star formation}
\label{sec:eagle_star_form}

Star formation is implemented following \cite{Schaye2008}, adopting the metallicity dependent density threshold of \cite{schaye2004}. It utilises the observed Kennicutt-Schmidt star formation law \citep{Kennicutt1998} but reformulated in terms of pressure to avoid dependence on the equation of state. This modified observed Kennicutt-Schmidt star formation law is implemented stochastically assuming a Chabrier initial mass function (IMF) \cite{chabrier_galactic_2003}.

Due to resolution limitations, it is necessary to invoke a star formation density threshold which defines the minimum density at which a stellar particle can be formed, this is defined as
\begin{equation}
    n^*_{\mathrm{H}}(Z) = 10^{-1} [\mathrm{cm}^{-3}] \left(\frac{Z}{0.002}\right)^{-0.64}
    \label{eq:star_thresh}
\end{equation}
where $n_H=10^{-1}$ cm$^{-3}$ is the critical volume density and $Z$ is the gas metallicity. For low metallicities this diverges, so an upper limit of $n^*_{\mathrm{H}}=10\ \mathrm{cm}^{-3}$ is imposed. Furthermore, a gas density threshold of 57.7 times the cosmic mean is implemented to prevent star formation in low overdensity gas at very high redshift.

\subsubsection{Stellar feedback}
\label{sec:eagle_star_fb}

Stars, particularly those that are massive and short lived, interact with the ISM by injecting energy via stellar winds, radiation and supernovae. In the \eagle\ model these energetic events, or stellar feedback events, are implemented following the stochastic thermal feedback method proposed by \cite{DallaVecchia2012}. In this implementation, the temperature increment ($\Delta T$) of a heated gas particle is specified explicitly and the contribution to a neighbouring gas particle is defined probabilistically. 

Stellar feedback happens only once: when a stellar particle reaches an age of 30 Myrs, a neighbouring SPH particle can be heated with a probability dependent on the fraction of the total amount of energy from core collapse supernovae per unit stellar mass that is injected. On average,
\begin{equation}
    f_\mathrm{th}= f_\mathrm{th, min} + \frac{f_\mathrm{th, max} - f_\mathrm{th, min}}{1 + \left(\frac{Z}{0.1 Z_\odot}\right)^{n_Z}\left(\frac{n_{\mathrm{H}, \mathrm{birth}}}{n_{\mathrm{H},0}}\right)^{-n_n}},
    \label{eq:fth}
\end{equation}
where $f_\mathrm{th, min}$ and $f_\mathrm{th, max}$ are asymptotes which can be chosen to tune stellar feedback in the model. The inclusion of the metallicity term, where $Z_\odot=0.0127$ is the solar metallicity and $n_Z=2 / \ln10$, accounts for the metallicity dependence of thermal losses in the ISM due to metal-line cooling. The inclusion of the density term, where $n_{\mathrm{H}, \mathrm{birth}}$ is the stellar birth density (the density inherited by the stellar particle from their parent gas particle), $n_n=n_Z=2/\ln10$, and the density pivot $n_{\mathrm{H},0}=0.67$ cm$^{-3}$, helps mitigate the inefficiency of feedback in highly enriched dense environments, such as compact galaxy cores. For low metallicities and high densities, $f_\mathrm{th}$ asymptotes to $f_\mathrm{th, max}$ (i.e stellar feedback is maximal), while at high metallicities and low densities $f_\mathrm{th}$ asymptotes to $f_\mathrm{th, min}$ (i.e stellar feedback is minimal). In the fiducial EAGLE models $f_\mathrm{th, max}=3$ and $f_\mathrm{th, min}=0.3$. \cite{crain_eagle_2015} showed that the choice of $f_\mathrm{th, max}$ is the more influential of the two, and choosing a value larger than unity better reproduces  galaxy stellar mass functions at low stellar masses.

\subsection{Structure Finding}
\label{sec:struct_find}

We follow the same structure extraction method used in the \eagle\ project and all previous \flares\ papers, explained in \cite{Mcalpine_data}. 
\begin{itemize}
    \item Dark Matter overdensities are identified using a Friends-Of-Friends (FOF) algorithm with a linking length of $\ell=0.2\bar{x}$, where $\bar{x}$ is the mean inter-particle separation.
    \item Baryonic particle species are then assigned to the halo containing their nearest dark matter neighbour.
    \item The FOF groups of both baryonic and dark matter particles are then refined by the \textsc{Subfind} algorithm \citep{springel_populating_2001, Dolag2009} to produce substructures (galaxies).
\end{itemize}

To refine FOF groups into galaxies, \textsc{Subfind} finds saddle points in the density field of a group. In pathological cases this can lead to undesirable splitting of genuine galaxies with regions of extreme density. These can then lead to spurious galaxies which are often mainly comprised of a single particle type. Although these pathological cases make up $<0.1\%$ of all galaxies with $M_\star>10^8$ M$_\odot$ at $z=5$, we nonetheless undo the erroneous splitting. To do so, we identify galaxies with no contributions in stellar, gas or dark matter components and recombine them into their nearest parent `central' substructure from which they were excised, if there is such a substructure within 30 pkpc.

In rare instances, tidal stripping can cause diffuse populations of particles at large radii. These separated populations can have extreme effects on derived galaxy properties. To mitigate the effect of this, we only consider particles within a 30 pkpc aperture centred on the galaxy's centre of potential, in line with previous \eagle\ and \flares\ projects. All galaxy properties presented in this work are derived using this aperture unless explicitly stated otherwise.

\section{Galaxy Half Mass Radii}
\label{sec:hmr}

In this section, we present 3-dimensional size-mass relations using half mass radii ($R_{1/2}$) derived from the particle distribution. The half mass radius is the radius of a sphere centred on the centre of potential enclosing half the total mass (of a particular particle species) in a 30 pkpc aperture.
Throughout this work, we only present half mass radii for galaxies with at least 100 stellar particles within the 30 pkpc aperture ($N_\star>100$), unless stated otherwise. This translates to an approximate stellar mass threshold of $M_\star \sim 10^{8} \mathrm{ M}_\odot$ and ensures the stellar distribution is well sampled and the derived half mass radii are robust. 

\subsection{Stellar Half Mass Radii}

\begin{figure}
	\includegraphics[width=\linewidth]{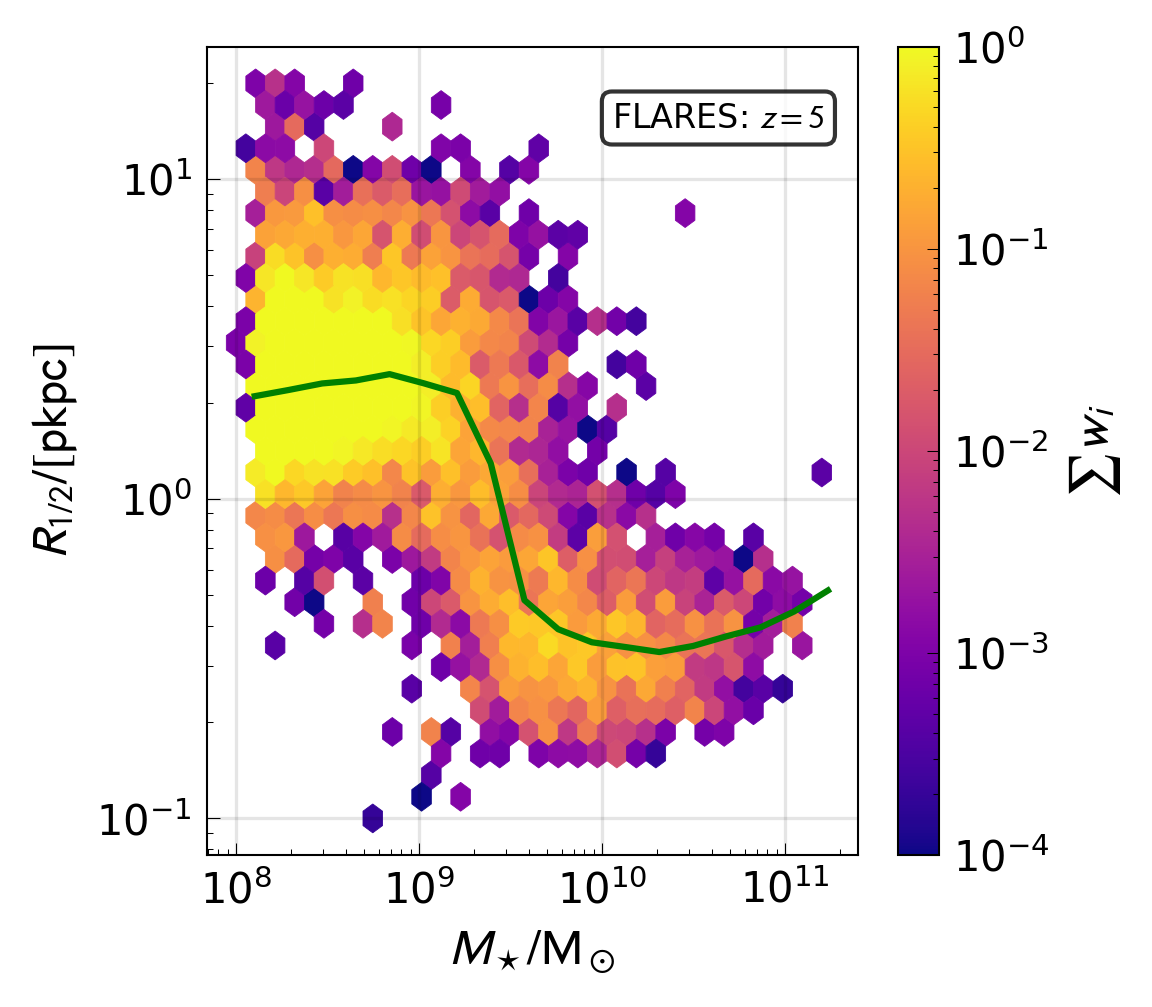}
	\includegraphics[width=\linewidth]{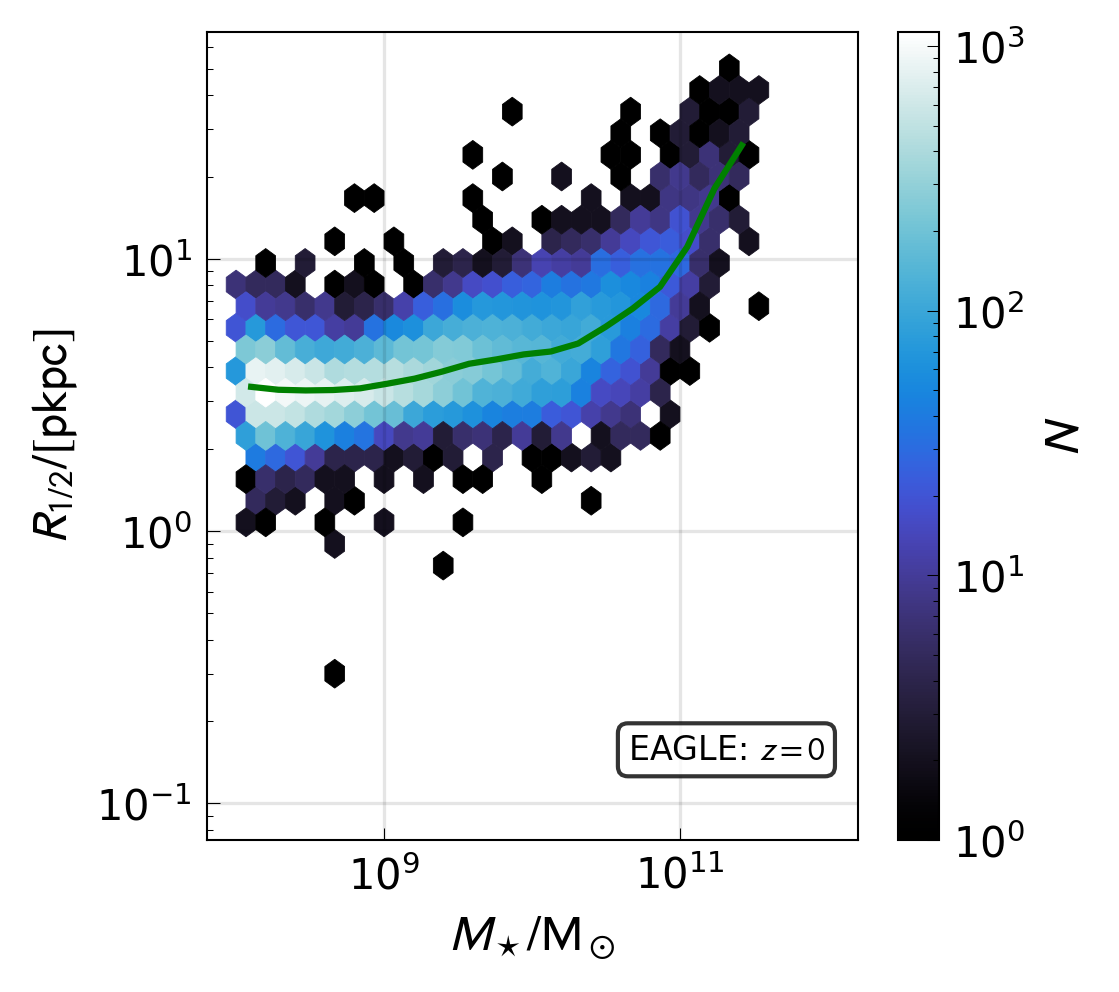}
    \caption{Upper panel: The stellar size-mass relation at $z=5$ for all galaxies in \flares\ with $N_\star>100$. The hexbins are coloured by the weighted number density derived using the \flares\ weighting scheme. Lower panel: The stellar size-mass relation at $z=0$ for all galaxies in \eagle-REF with $N_\star>100$. The hexbins are coloured by the number of galaxies in a bin. In both panels, the green curves show the 50th percentile of the distribution.}
    \label{fig:star_hmr}
\end{figure}

\fig{star_hmr} shows the stellar size-mass relation at $z=5$ from \flares\ (top panel) and $z=0$ from \eagle\ (bottom panel). The \flares\ galaxies in the upper panel are statistically weighted using the overdensity of their resimulation region to ensure the resimulated sample is representative of the population. For a detailed description of this weighting, we direct the reader to \cite{Lovell2021}.
At low redshift, the relation has a positive slope, as demonstrated in \cite{Furlong_2017}. However, in \flares\ we find a negatively sloped and bi-modal stellar size-mass relation. Although not shown here, the bi-modality and negative trend are evident for all snapshots at $z\geq5$ \citep{Roper22} and agree with the findings of other works \citep{Wu_20, Marshall21, Popping2021}. For galaxies in \flares\ at $z=5$ to become comparable in size to those found in \eagle\ at $z=0$ they will have to become at least ten, or for the most massive galaxies at least 50, times larger.
This bi-modality implies two scenarios: there are two distinct populations of galaxies with separate evolution tracks, or galaxies start as diffuse low mass systems which later evolve to become compact high mass systems. We investigate which scenario is applicable in \sec{size_mass_evo}.

\subsection{Gas Half Mass Radii}
\label{sec:gas_hmr}

\begin{figure}
	\includegraphics[width=\linewidth]{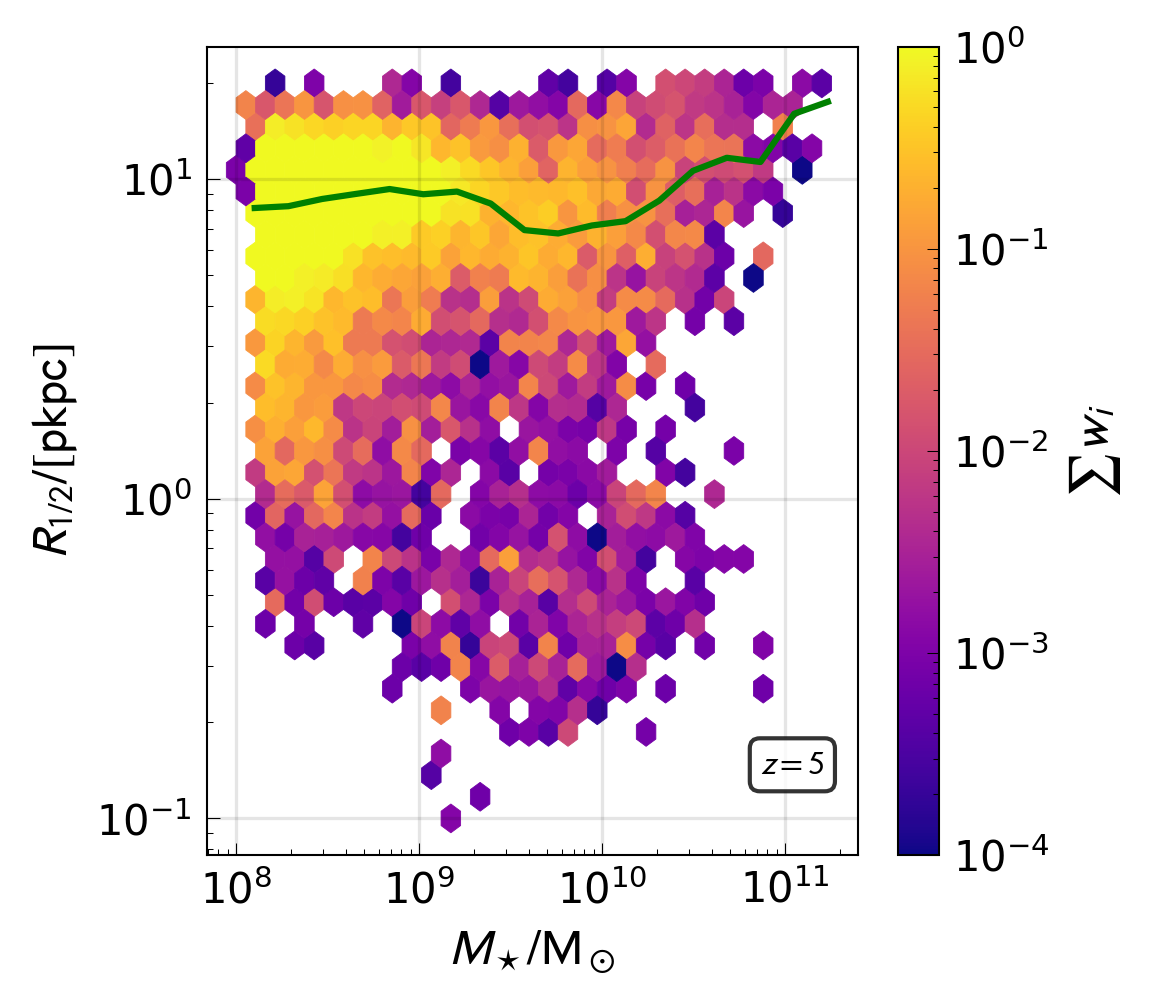}
    \caption{The gas size-stellar mass relation at $z=5$ for all galaxies in \flares\ with $N_\star>100$. The hexbins are coloured by the weighted number density derived using the \flares\ weighting scheme and the solid green line shows the 50th percentile of the distribution.}
    \label{fig:gas_hmr}
\end{figure}

If high redshift stellar distributions are compact, then one might expect that the gas from which they form should also be compact. \fig{gas_hmr} shows the gas size-stellar mass relation at $z=5$ for all galaxies in \flares\ with $N_\star>100$. In contrast to the stellar size-mass relation, we find the relation between gas size and stellar mass is constant with a large scatter to small gas radii at fixed stellar mass. For galaxies with $M_\star/\mathrm{M}_\odot > 10^9$ there again appears to be a bi-modality in gas size. Note, however, that this bi-modality is different to that present in \fig{star_hmr} where galaxies with $M_\star/\mathrm{M}_\odot > 10^{10}$ populate only the compact population.

\begin{figure}
	\includegraphics[width=\linewidth]{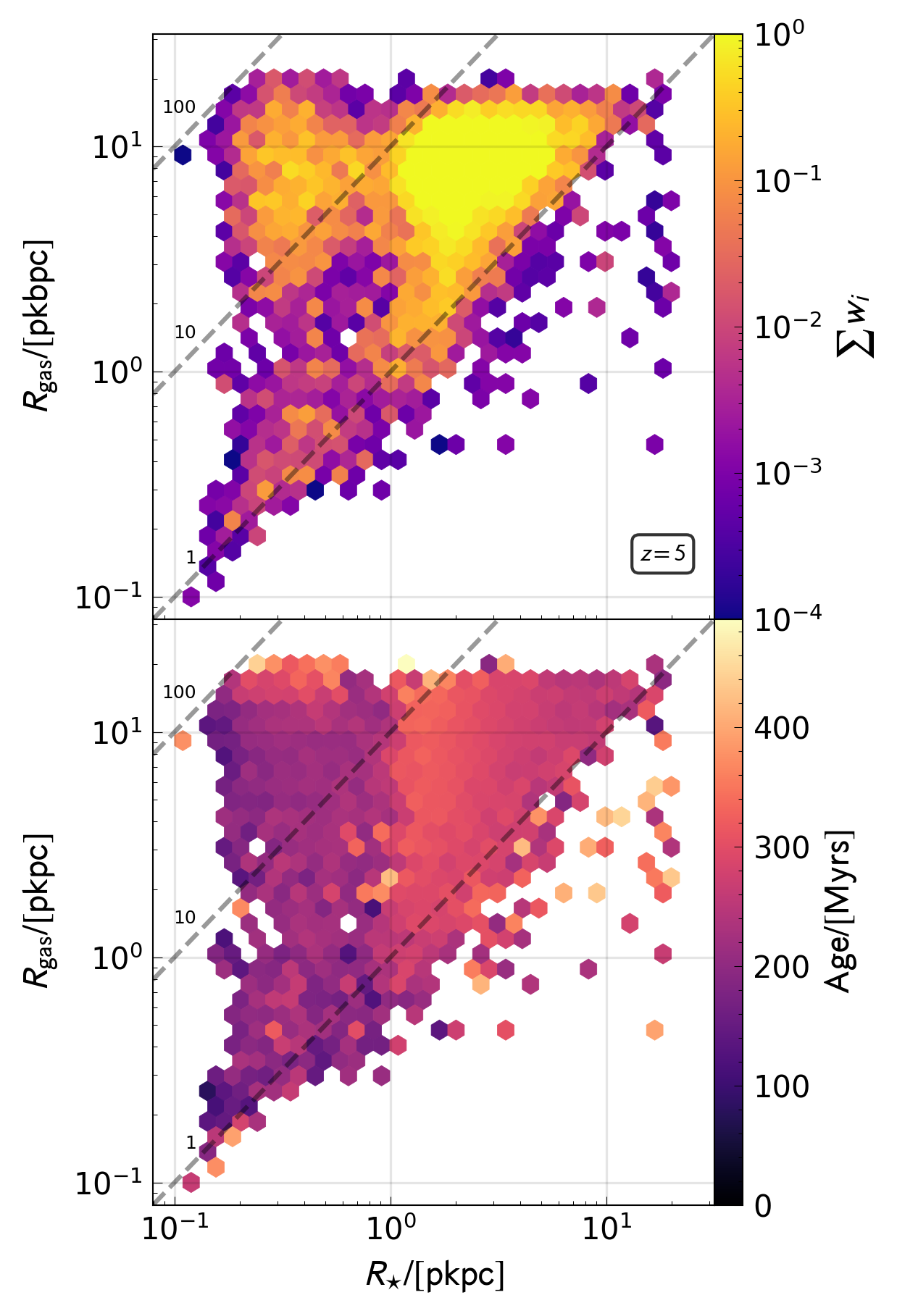}
    \caption{A comparison of stellar half mass radii and gas half mass radii for all galaxies in \flares\ at $z=5$ with more than 100 stellar particles. The hexbins in the upper panel show the number density weighted using the \flares\ weighting scheme. The lower panel shows the mean galaxy age in each hexbin, as defined by the initial stellar mass weighted age. The dashed lines show $R_{\rm gas} / R_\star$ in powers of ten to aid interpretation.}
    \label{fig:gas_hmr_comp}
\end{figure}

To explicitly compare stellar distributions to gas distributions we compare in the upper panel of \fig{gas_hmr_comp} the gas half mass radii, including all gas present in the galaxy, to stellar half mass radii at $z=5$ for all galaxies in \flares\ with $N_\star>100$. From this, we can see that compact stellar distributions are associated with extended non-star forming gas distributions in most cases. These extended gas distributions can be as much as $\sim100\times$ larger than the stellar component. We can now see the bi-modality in \fig{gas_hmr} is due to massive galaxies which are not associated with extended gas distributions but instead have compact gas components comparable to their stellar component.

One possible explanation for the difference between the gas distributions of compact galaxies could be that the galaxies with compact gas distributions are newly formed, and are yet to accrete the extended gas distributions of their gas enshrouded brethren. In the lower panel of \fig{gas_hmr_comp} we probe this by colouring the hexbins by the mean age of galaxies in each hexbin, where we have defined a galaxy's age as the initial stellar mass weighted average of stellar particle age. This shows no discernible difference between the age of galaxies with compact or extended gas distributions. We can therefore confidently throw out the accretion hypothesis.

\begin{figure}
    \centering
    \includegraphics[width=\linewidth]{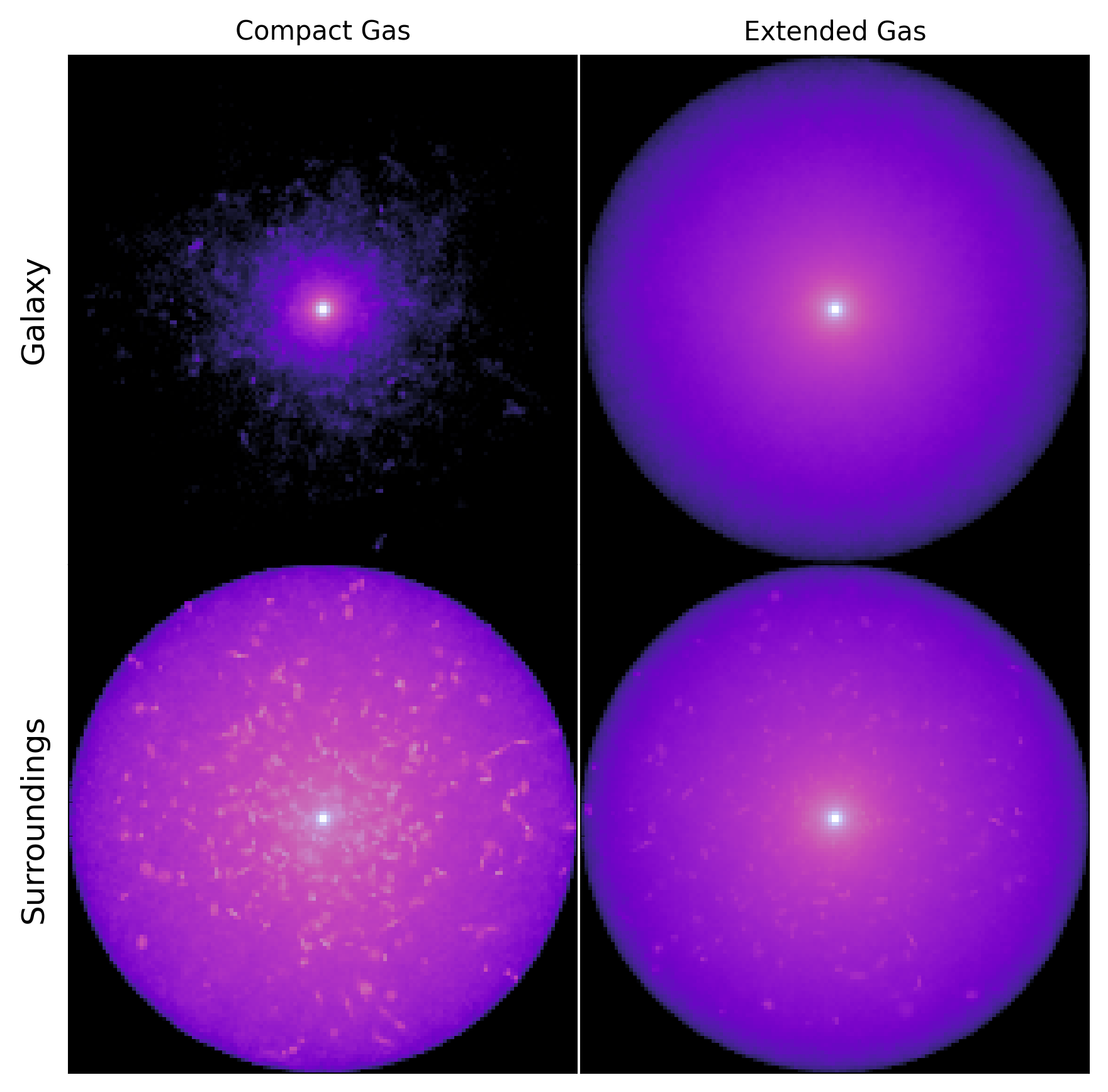}
    \caption{Stacked images comparing the gas distributions of galaxies in all \flares\ regions at $z=5$ with compact stellar distributions and extended or compact gas distributions. The left hand column shows galaxies with compact gas distributions, where $R_{1/2, \mathrm{gas}}< 1$ pkpc and $R_{1/2, \star} < 1$ pkpc, while the right hand column shows galaxies with extended gas distributions, where $R_{1/2, \mathrm{gas}}>= 1$ pkpc and $R_{1/2, \star}< 1$ pkpc. The top row shows the particles identified by SUBFIND to belong to the galaxy while the bottom row shows all particles within 30 pkpc of the galaxy's centre of potential. Each image is 60 pkpc in width, the pixel resolution is the softening length of the simulation and gas particles have not been smoothed over their kernels.}
    \label{fig:gas_img}
\end{figure}

A possible numerical explanation for the split between compact and diffuse gas distributions is that the halo finder has erroneously split galaxies apart from the extended gas distributions surrounding them. To ascertain if this is the case we show stacked images of the gas distribution in \fig{gas_img} split by gas half mass radius (left column: $R_{1/2, \mathrm{gas}}< 1\,$pkpc, right column: $R_{1/2, \mathrm{gas}}> 1\,$pkpc). Each panel has a width of 60 pkpc and a pixel resolution equal to the softening length of the simulation. The top row shows the difference between the compact and extended gas distributions as defined by SUBFIND, whereas the bottom row shows the stacks but includes all stellar particles within 30 pkpc of each galaxy's centre (defined by the centre of potential). 
If the compact gas population were the result of a misidentification by the halo finder, we would expect to see similarly extended profiles in both panels in the bottom row. However, the compact gas distribution stack is noticeably less extended. So we can conclude this bi-modality is not the result of numerical effects in structure definition.

These compact stellar distributions associated with compact gas distributions are therefore legitimate structures. However, of the 425 galaxies with compact gas distributions, 418 are satellites, leaving only 7 central galaxies. These satellites have likely undergone tidal stripping and are unlikely to persist to low redshift in their current compact gas-poor form without undergoing mergers. It is worth noting that although the majority of compact gas distributions are associated with satellites, it is not true that the majority of satellites have compact gas distributions. 
On the other hand, the centrals with compact stellar and gas distributions will not be able to accrete gas and continue forming stars in the near future, and could lead to gas-poor passive compact sources in the future. These could be indicative of one possible evolution track that leads to the quiescent red nuggets seen in observations at lower redshifts \citep[e.g.][]{van_Dokkum_08, Damjanov_09, Taylor_10}.

\section{Compact Galaxy Formation}
\label{sec:form}

In this section we probe the formation of massive compact galaxies in the \eagle\ model. We showed in \cite{Roper22} that the compact galaxies driving the negative slope of the size-mass relation (see \fig{star_hmr}) produce a positively sloped size-luminosity relation due to the effects of dust. This size-luminosity relation is in good agreement with observations. We now explore the mechanism driving the formation of these massive compact galaxies, and explore why they only exist above a certain stellar mass threshold ($M_\star > 10^{9} {\rm M}_\odot$). 

To probe the size and mass evolution of individual galaxies we employ the MErger Graph Algorithm \citep[MEGA,][]{Roper_20}\footnote{MEGA has been updated to construct merger graphs for hydrodynamical simulations, in addition to purely $N$-body codes. It can also now use halos defined by other algorithms rather than those produced with its own phase space halo finder. Note, however, that merger graphs generated from other input halo catalogues will not benefit from the temporally motivated approach utilised in MEGA.}
to construct merger graphs from the SUBFIND halo catalogue. Throughout the following sections we define:

\begin{itemize}
    \item A progenitor of a galaxy as any galaxy in the previous snapshot which contributed at least 10 particles (of any type) to the galaxy.
    \item The main progenitor as that which contributed the most particles (of any type) to the galaxy.
    \item The main branch as the chain of main progenitors starting with a galaxy and ending with it's earliest main progenitor.
    \item The root as the first galaxy in the main branch at $z=5$.
\end{itemize}

For the purposes of graph construction, all SUBFIND halos are included down to a particle count of 20, unlike the sample used to probe morphological quantities which is limited to those with $N_\star>100$. Throughout all subsequent plots, all galaxies considered in snapshot B obey the $N_\star>100$ threshold but progenitors are presented without applying any limit. It is worth noting that the cadence of \flares\ snapshots is sub-optimal for merger graph construction. However, this does not affect the robustness of main branch definitions and only poses a problem if the merger graphs were to be used for semi-analytical modelling or assessing the impact of mergers. 

\subsection{Evolution across the size-mass relation}
\label{sec:size_mass_evo}

\begin{figure*}
	\includegraphics[width=\linewidth]{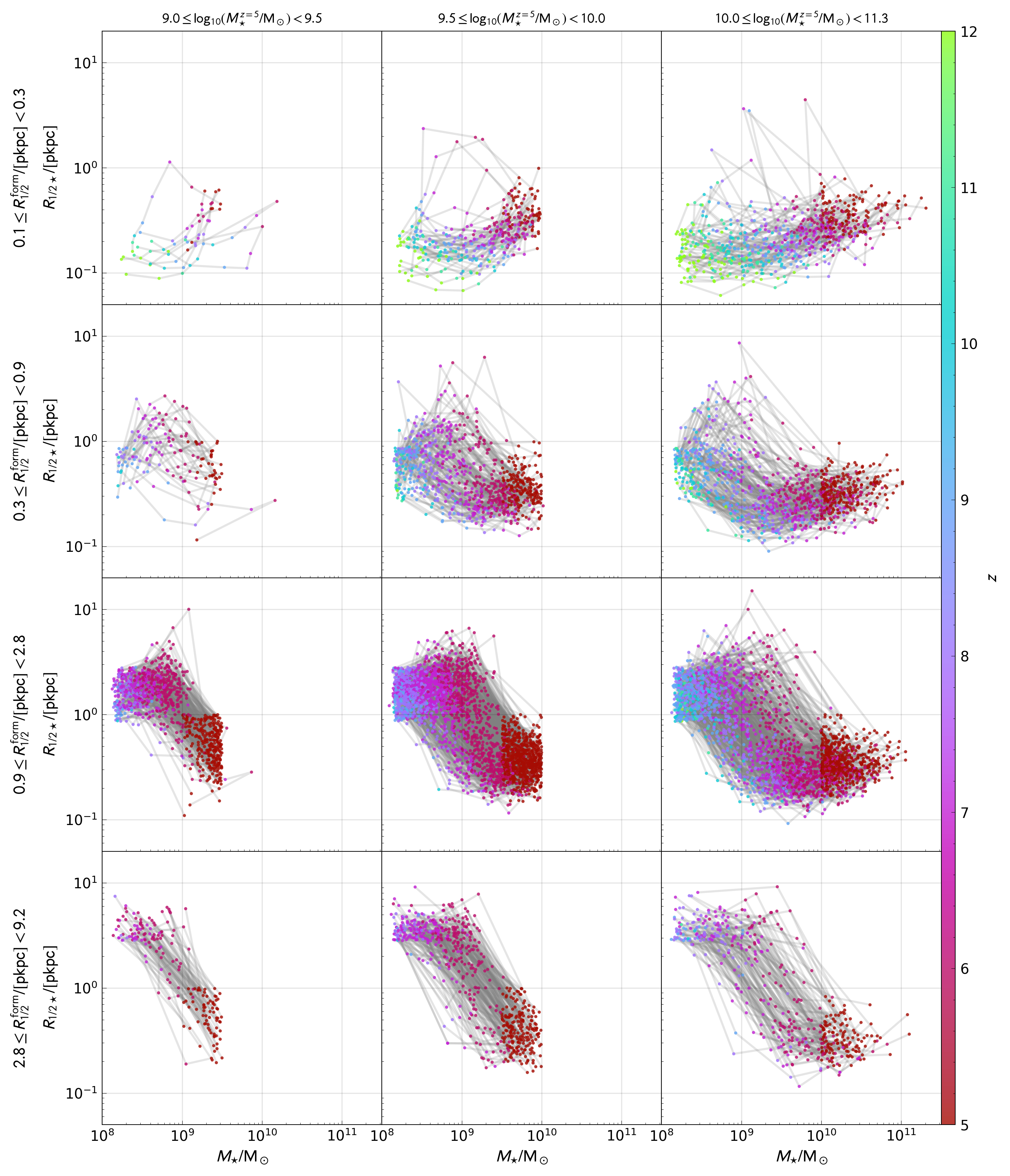}
    \caption{The stellar size-mass relation evolution for galaxies which have compact stellar distributions at $z=5$ ($R_{1/2}<1$ pkpc). The galaxies are divided into size and mass bins, where the columns are binned by the stellar mass of the galaxy at $z=5$, and the rows are binned by the initial stellar half mass radius of a galaxy when it enters the sample. Points are coloured by their redshift.}
    \label{fig:hmr_evo}
\end{figure*}

To identify the evolution of galaxies causing the bi-modality in the $z=5$ size-mass relation (\fig{star_hmr}) we present the evolution across the size-mass relation of galaxies with stellar half mass radii $<1$ pkpc at $z=5$ in \fig{hmr_evo}. To do so we plot each individual main branch for these galaxies rooted at $z=5$, splitting into initial size and final mass bins to aid interpretation. These main branches yield two evolution paths for compact galaxies at $z=5$. 

In the first formation path, all progenitors along the main branch are compact including their earliest low mass progenitor. Although these galaxies remain compact throughout their evolution, they exhibit a positive size evolution. We denote this formation path the compact formation path. 

In the second, more dominant formation path, galaxies begin as diffuse low mass galaxies which then decrease in size between adjacent snapshots. We denote this formation path the diffuse formation path. From this, we can conclude that in reality, both the formation scenarios proposed in \sec{hmr} are applicable, but each is applicable in different redshift regimes.

The population with compact formation paths are galaxies forming at the earliest times ($z>10$). Gas at these redshifts has undergone minimal enrichment requiring a higher density threshold to form stellar particles (see \sec{eagle_star_form}). The higher density threshold leads to star formation delayed until gas has collapsed further, yielding concentrated starbursts and thus initial compact stellar half mass radii. 

Once the gas distribution has been sufficiently enriched star formation can proceed at lower densities, giving rise to galaxies with the diffuse formation path. These galaxies enter the catalogue at $z\leq10$ as diffuse systems before undergoing a mechanism causing a decrease in size, yielding a compact galaxy. Compact galaxies in the lowest final mass bin (left hand column) undergo this transition between $z=6$ and $z=5$. Those in the higher final mass bins become compact at earlier redshifts, with many in the highest mass bin (right hand column) exhibiting a positive size evolution after joining the compact distribution. This relation between final mass and transition redshift shows that not only is this an ongoing process at $z=5$ but also that the size transition takes place in a particular stellar mass regime. Once compact these galaxies continue to form stellar mass and increase in size as they do so. 

In \fig{hmr_evo} there are a small number of outliers with pathological increases in size and in some cases decreases in stellar mass, such as those in the top right panel with half mass radii $>1$ pkpc. These are examples of structure finding issues discussed in detail in \cite{SMT}. In cases such as these interacting galaxies have either been temporarily amalgamated leading to increased sizes and masses or mass associated with a galaxy has been temporarily misassigned to a neighbouring galaxy. 

\subsection{Changes in size}

\begin{figure*}
	\includegraphics[width=\linewidth]{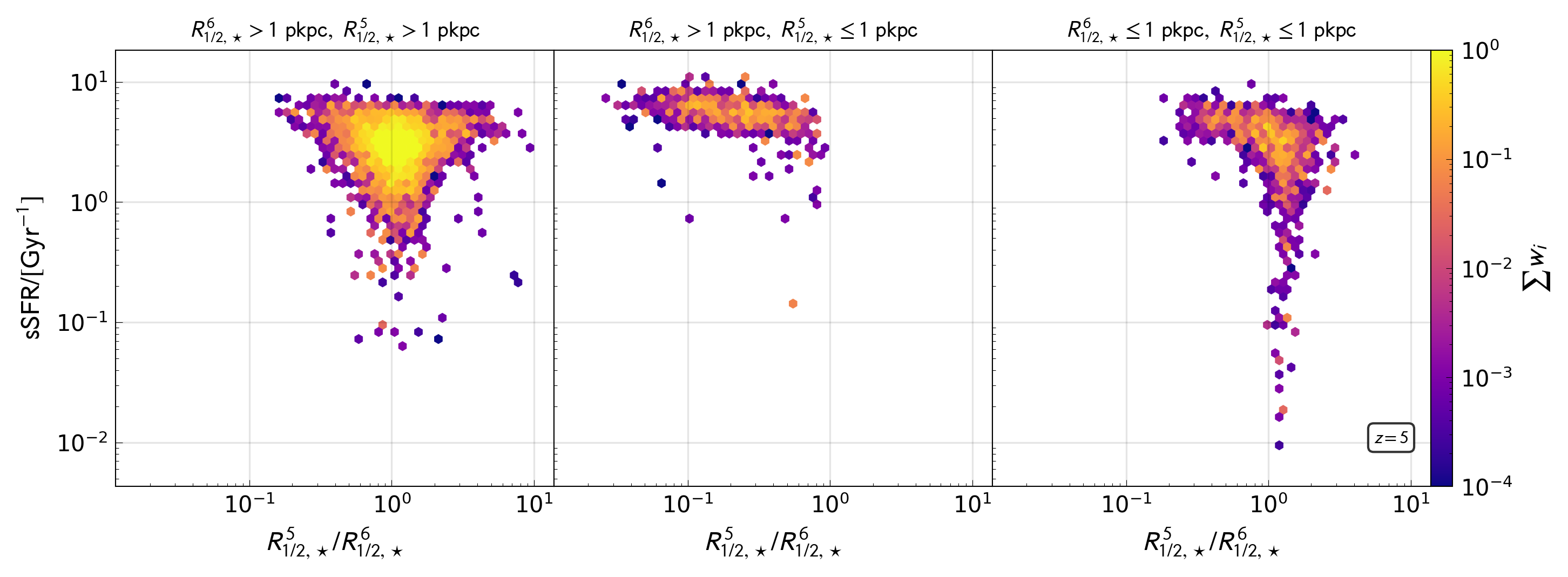}
    \caption{The specific star formation rate of galaxies as a function of the ratio between stellar half mass radii at $z=5$ and their main progenitors' at $z=6$. Each panel shows galaxies undergoing different phases of size evolution, left to right: galaxies which were diffuse at $z=6$ and remain diffuse at $z=5$; galaxies which were diffuse at $z=6$ and are compact at $z=5$; and galaxies which were compact at $z=6$ and remain compact at $z=5$. Hexbins are coloured by weighted number density using the \flares\ region weighting scheme.}
    \label{fig:hmr_ssfr}
\end{figure*}

The diffuse formation path in \fig{hmr_evo} contains a transition between the diffuse and compact regimes evident in the size-mass relation shown in \fig{star_hmr}. This transition is governed by a mechanism taking place at a particular stellar mass. In this section we investigate the mechanism driving this transition from diffuse stellar components to compact stellar components.

In \fig{hmr_ssfr} we show the specific star formation rate (sSFR) of galaxies at $z=5$ as a function of the change in stellar size between $z=6$ and $z=5$, defined as a ratio between a galaxy's size and their main progenitor's size at $z=6$. We define the sSFR using all stellar particles formed within 100 Myr in a 30 pkpc aperture. A size ratio $>1$ suggests a galaxy has increased in size while a size ratio $<1$ suggests a galaxy has decreased in size.

We can see that galaxies which were diffuse at $z=6$ and remain diffuse at $z=5$ (left hand panel) have no clear trend between their sSFR and change in size. Galaxies which remain compact between $z=6$ and $z=5$ (right hand panel) show both decreases ($R^{5}_{1/2} / R^{6}_{1/2} < 1$) and increases ($R^{5}_{1/2} / R^{6}_{1/2} > 1$) in size, the majority of the latter have lower sSFRs than the former. In contrast, galaxies which transition from diffuse to compact (central panel) are highly star forming with a subtle negative trend between sSFR and $R^{5}_{1/2} / R^{6}_{1/2}$. The tail of passive galaxies (sSFR $<0.1$) exhibit small increases in size ($R^{5}_{1/2} / R^{6}_{1/2} \sim 1.2$). This size increase is likely due to stellar migration, where stellar particles move into larger orbits, as found at low redshift in \cite{Furlong_2017}. These passive galaxies are explored in detail in \cite{Lovell22}. 

All galaxies undergoing the transition between diffuse and compact regimes have high sSFRs, while galaxies not undergoing this transition can exhibit a range of behaviours. From this we can conclude that the mechanism driving this transition is centrally concentrated star formation rather than stellar migration.

\begin{figure*}
	\includegraphics[width=\linewidth]{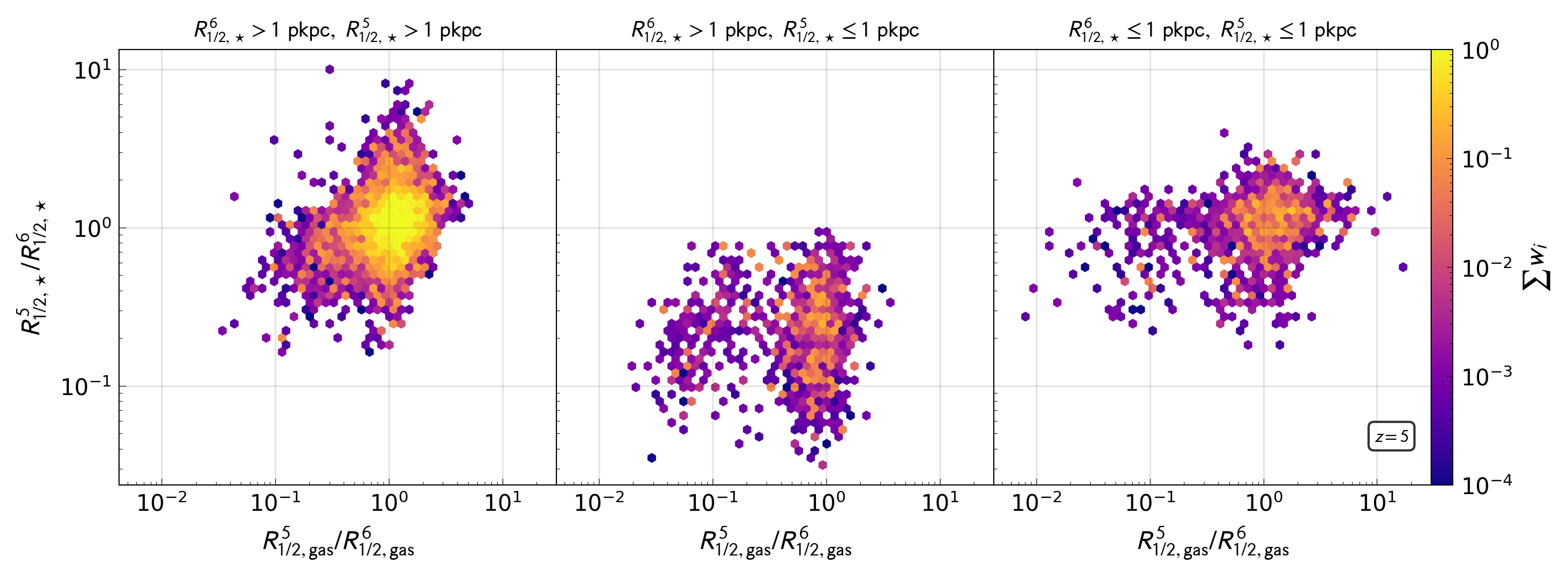}
    \caption{Change in stellar half mass radii vs change in gas half mass radii between galaxies at $z=5$ and their main progenitors at $z=6$. Each panel shows galaxies undergoing different phases of size evolution, left to right: galaxies which were diffuse at $z=6$ and remain diffuse at $z=5$, galaxies which were diffuse at $z=6$ and are compact at $z=5$, and galaxies which were compact at $z=6$ and remain compact at $z=5$. Hexbins are coloured by weighted number density using the \flares\ region weighting scheme.}
    \label{fig:hmr_star_gas_comp}
\end{figure*}

We now probe the changes in the gas distribution in \fig{hmr_star_gas_comp}. This shows the change in stellar half mass radius vs the change in gas half mass radius at $z=5$. In each panel, we see little correlation between the change in stellar size and the change in gas size. A small number of galaxies' gas distributions shrink by a factor of $\sim10$, but these galaxies show no trends in change in stellar size. These shrinking gas distributions are those associated with the compact stellar distributions in \fig{gas_hmr_comp}, explicitly showing the change in gas half mass radius due to tidal stripping of the satellites in that population (as discussed in \sec{gas_hmr}). There is a subset of diffuse galaxies that remain diffuse (left hand panel) where a decrease in size of the stellar component is accompanied by a decrease in size of the gas component, this is driven by gas cooling leading to collapsing gas clouds and thus concentrated star formation causing a decrease in stellar size (we present more on this in \sec{cool_rad}).

The lack of gas half mass radius change in \fig{hmr_star_gas_comp} should come as no surprise given the extended nature of gas distributions detailed in \fig{gas_hmr_comp}. The vast majority of the gas distributions are spatially distinct from the regions of compact star formation and are thus unaffected by the energy injected by stellar feedback.

\subsubsection{The role of stellar feedback}
\label{sec:star_fb}

\begin{figure*}
        \includegraphics[width=\linewidth]{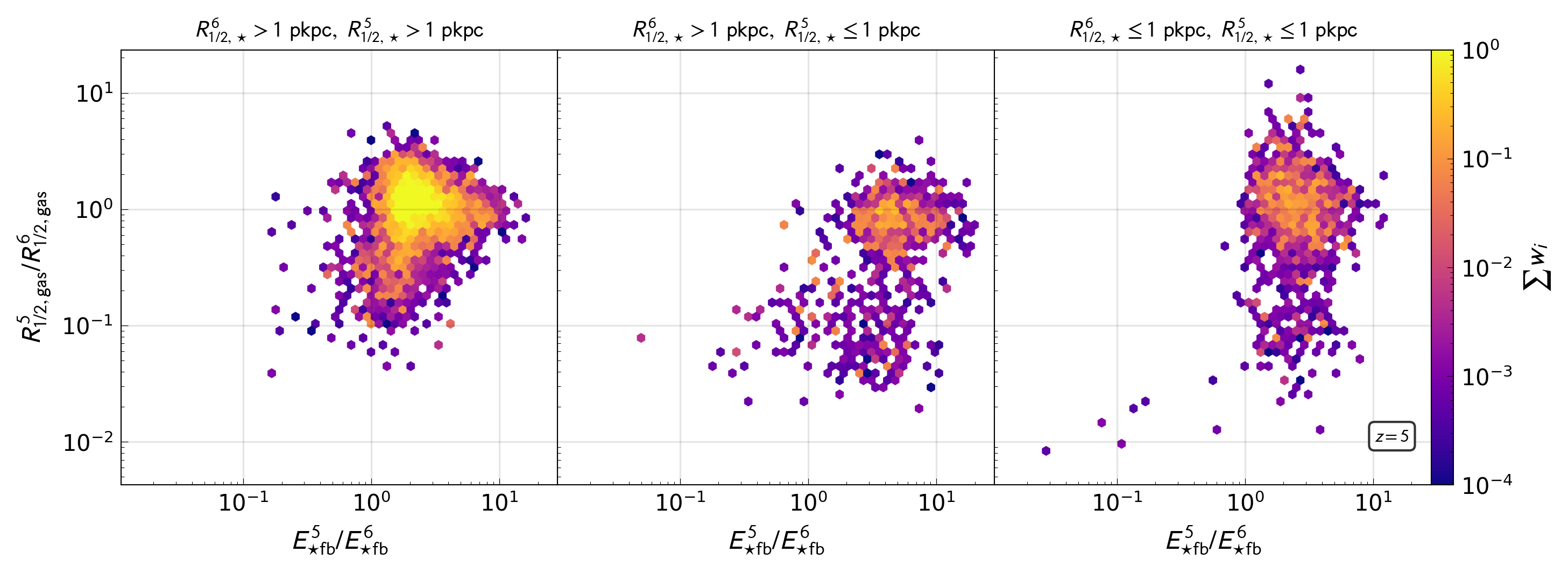}    
	\includegraphics[width=\linewidth]{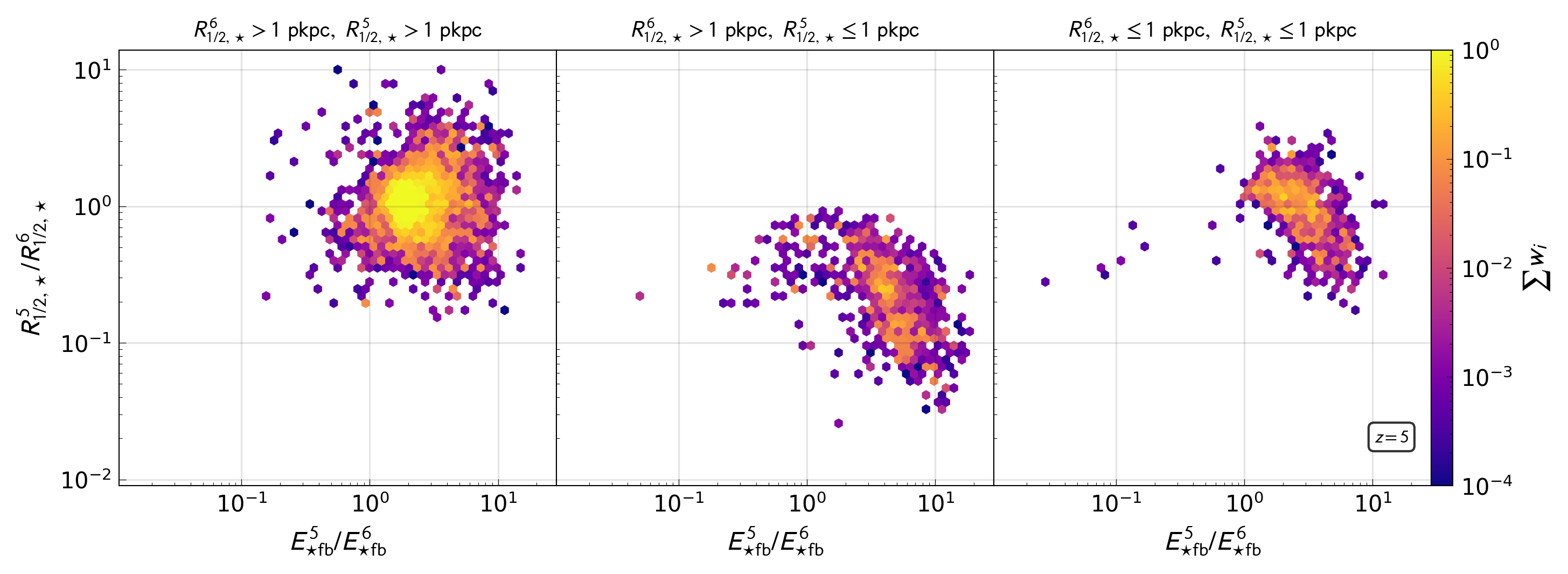}
    \caption{Upper panel: Gas half mass radii ratios as a function of integrated stellar feedback ratios. Lower panel: Stellar half mass radii ratios as a function of integrated stellar feedback ratios. In both rows, each panel shows galaxies undergoing different phases of size evolution, left to right: galaxies which were diffuse at $z=6$ and remain diffuse at $z=5$, galaxies which were diffuse at $z=6$ and are compact at $z=5$, and galaxies which were compact at $z=6$ and remain compact at $z=5$. The integrated feedback energy is defined by \eq{efb}, and the hexbins are coloured by weighted number density using the \flares\ region weighting scheme.}
    \label{fig:fb_star_gas_comp}
\end{figure*}

It was shown in \cite{crain_eagle_2015} that stellar feedback in the \eagle\ model is inefficient at high redshift, leading to high sSFR rates in regions of high density. However, the large localised SFRs necessary to transition from the diffuse regime to the compact regime shown in \fig{hmr_ssfr} will dump disproportionate amounts of thermal energy into their dense surroundings. Here we probe the effect of stellar feedback during the size transition process and thus further probe the efficiency of stellar feedback in this regime.

In \fig{fb_star_gas_comp} we show the effect of stellar feedback on both the stellar distributions and gas distributions. Here we plot the change in half mass radius as a function of the change in integrated stellar feedback energy between compact galaxies at $z=5$ and their progenitors at $z=6$. We define the integrated stellar feedback energy following \cite{Davies_20} (Eq. 4 therein) as 
\begin{equation}
\label{eq:efb}
    E_{\star\mathrm{fb}} = \sum_{i=0}^{i=N_\star}1.74\times10^{49} \mathrm{erg} \left(\frac{m_{\star,\mathrm{init},i}}{1\mathrm{M}_\odot}\right)f_{\mathrm{th},i},
\end{equation}
where $m_{\star,\mathrm{init},i}$ is the initial mass of a stellar particle and $f_{\mathrm{th}, i}$ is the feedback fraction of individual stellar particles given by \eq{fth}. We obtain the integrated stellar feedback energy by summing over the contribution of all stellar particles in a galaxy within a 30 pkpc aperture. Note that, given this definition, the ratio $E^{5}_{\star\mathrm{fb}}/E^{6}_{\star\mathrm{fb}}$ can be less than unity if stellar particles are lost between $z=5$ and $z=6$. Instances where $E^{5}_{\star\mathrm{fb}}/E^{6}_{\star\mathrm{fb}}<1$ are either due to galaxy interactions or misidentification by the halo finder.

The upper row of \fig{fb_star_gas_comp} clearly shows little correlation between the change in integrated stellar feedback and the change in gas half mass radius. This is due to the aforementioned extent of the gas distribution showing that the effects of stellar feedback remain localised in the core and have no effect on the greater distribution. However, the same is not true for the stellar distribution. In the lower row of \fig{fb_star_gas_comp} we present the change in stellar half mass radius as a function of the change in integrated stellar feedback. In the central panel, containing galaxies undergoing the size transition, we find a clear trend where larger increases in $E_{\star\mathrm{fb}}$ yield larger decreases in stellar half mass radius. For galaxies remaining compact (right hand panel) we find a shallower trend at larger change in size ratios with significant scatter at fixed feedback energy ratio. Notably, even in extreme cases the amount of stellar feedback is still too inefficient to be capable of slowing down star formation. 

The galaxies which remain diffuse (left hand panel) exhibit no trend between change in size and integrated stellar feedback. In this regime not only are fewer stars forming, and therefore fewer stellar feedback events, but the gas is also significantly less dense yielding a lower $f_\mathrm{th}$. It is therefore unsurprising we see no trend in this regime.

\subsection{Dynamical effects}

Given the mechanism controlling the formation of compact galaxies detailed in the previous section, we now probe the dynamics and attempt to determine why the compact to diffuse transition takes place at $M_\star/\mathrm{M}_\odot\sim 10^9$. To do so, we define the total binding energy of a galaxy as
\begin{equation}
    E_{\mathrm{bind}}=G\sum_{i=0}^{i=N_\star}\sum_{j=i + 1}^{j=N_\star} \frac{m_i m_j}{\sqrt{r_{ij}^2 + \varepsilon^2}},
    \label{eq:ebind}
\end{equation}
where $G$ is the gravitational constant, $m_{i/j}$ are the masses of particles $i$ and $j$, $r_{ij}$ is the distance between particles $i$ and $j$, and $\varepsilon$ is the softening of the simulation in physical coordinates. We calculate this quantity including all particle species identified to be part of a galaxy.

\begin{figure}
	\includegraphics[width=\linewidth]{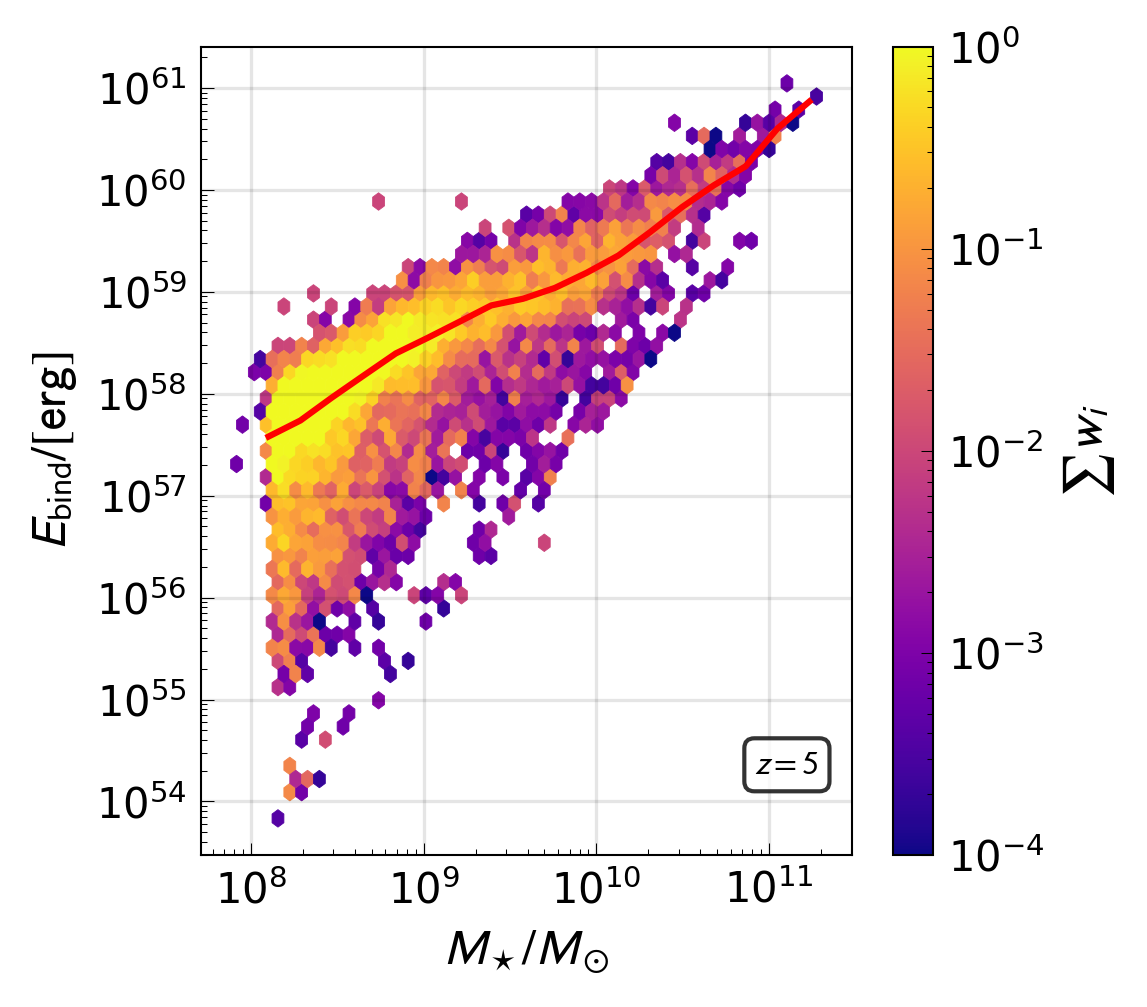}
	\includegraphics[width=\linewidth]{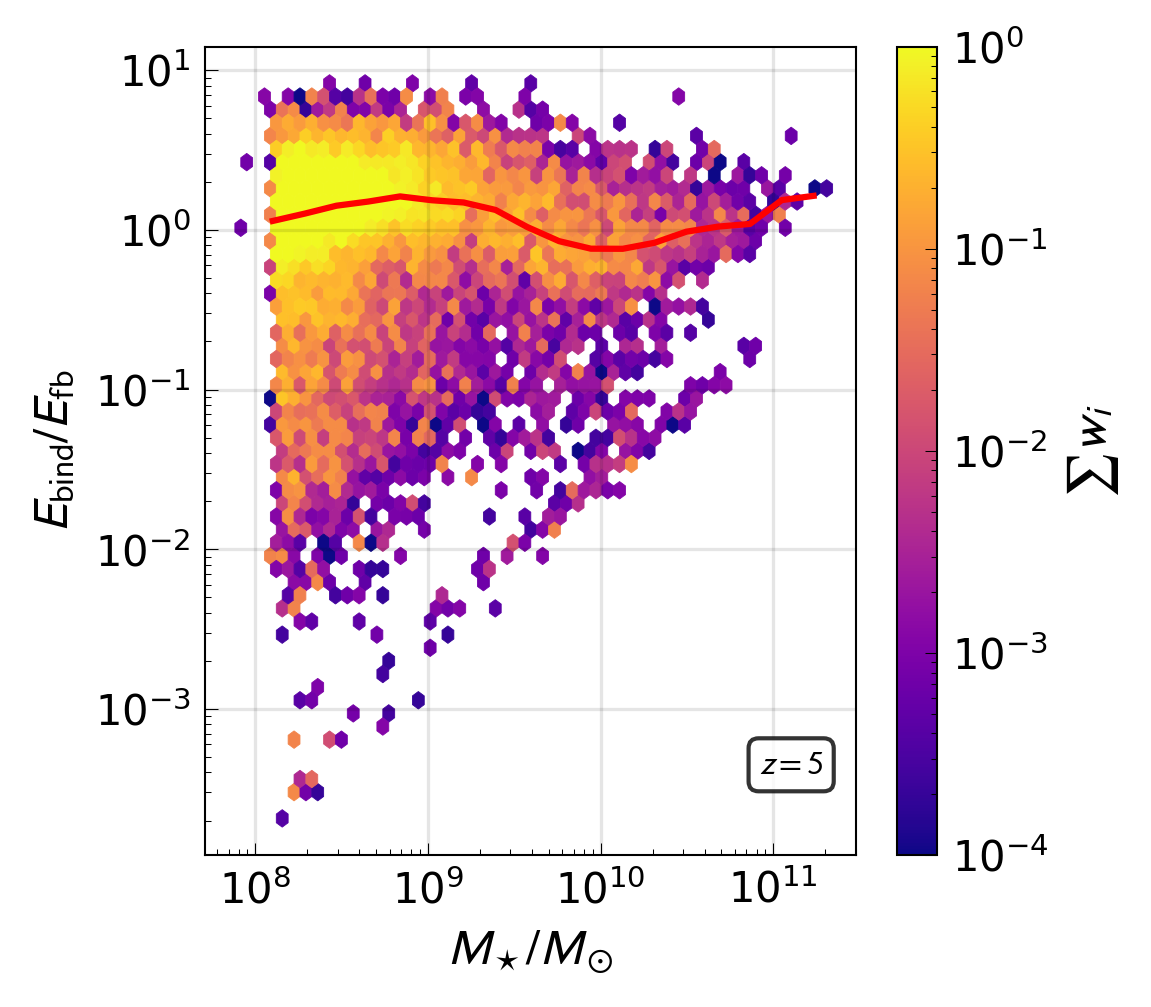}
    \caption{Upper panel: The gravitational binding energy as a function of stellar mass for all galaxies at $z=5$. Lower panel: The ratio between gravitational binding energy and the integrated stellar feedback energy as a function of stellar mass for all galaxies at $z=5$. In both panels the gravitational binding energy is defined by \Eq{ebind}, the integrated feedback energy is defined by \Eq{efb}, and the hexbins are coloured by weighted number density using the \flares\ region weighting scheme. The solid line shows the 50$^{\rm th}$ percentile of the distribution.}
    \label{fig:bind_mass}
\end{figure}

In the upper panel of \fig{bind_mass} we present the stellar mass dependence of the total binding energy (\eq{ebind}). There is a clear positive linear relation between total binding energy and stellar mass with a large scatter at low masses which decreases with increasing stellar mass. The binding energy for the massive galaxies at which we see the compact transition ($M_\star/\mathrm{M}_\odot\sim 10^9$) exhibits a small drop away from the linear trend of the lower mass galaxies. This can be attributed to the effects of stellar feedback in these highly star forming galaxies. Although stellar feedback is too inefficient and localised to strongly affect the morphology of a galaxy it nonetheless affects the gas distribution in the region where gravitational attraction between particles is greatest, and thus where the total binding energy is most sensitive to effects on the gas distribution. 

We further investigate the interplay between total binding energy and stellar feedback in the lower panel of \fig{bind_mass}, which shows the ratio between total binding energy and integrated stellar feedback (detailed in \sec{star_fb}) as a function of stellar mass. Here we can explicitly see the drop in total binding energy relative to integrated stellar feedback for the highly star forming compact galaxies with $M_\star/M_\odot > 10^9$. Note that these galaxies are nonetheless bound, as the injected thermal energy from stellar feedback is radiated away by the efficient cooling in star forming regions on short enough timescales.

From \fig{bind_mass} we can see no obvious features in the overall dynamics of the entire galaxy that enable the efficient localised star formation that drives the compactification of galaxies at $M_\star/M_\odot\sim 10^9$. We can thus conclude that this transition is not a dynamical mechanism affecting both baryonic and dark matter components of a galaxy.

\subsection{Cooling radii}
\label{sec:cool_rad}

To find the cause of the mass threshold at which the size transition takes place we need to probe the behaviour and distribution of star forming gas particles. To do this we employ a `cooling radius' ($R_{\mathrm{cool}}$). We define a `cooling radius' of a galaxy as the gas half mass radius weighted by gas density. This is effectively a measure of the size of the star forming gas distribution. We use this definition to account for variations in density within star forming regions and take into account gas particles which soon will be star forming.

\begin{figure}
	\includegraphics[width=\linewidth]{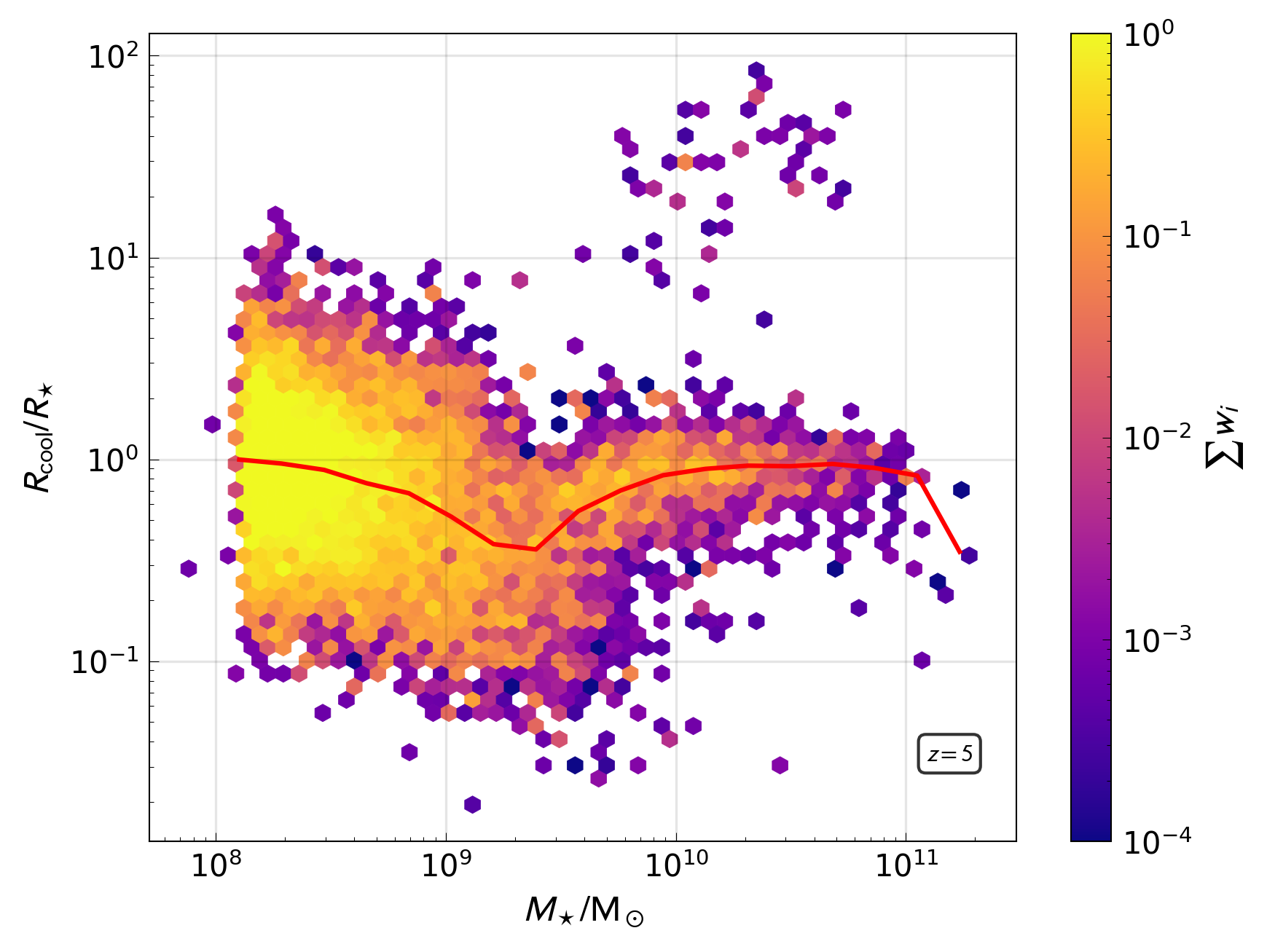}
    \caption{The ratio between gas density weighted gas half mass radii (cooling radii, $R_{\rm cool}$) and stellar half mass radii as a function of stellar mass at $z=5$. The solid line denotes the weighted 50th percentile of this distribution. Hexbins are coloured by number density weighted using the \flares\ weighting scheme. Here we have placed an additional cut on the sample such that included galaxies have both $N_\star > 100$ and $N_\mathrm{gas} > 100$ to ensure size measurements are robust.}
    \label{fig:cool_rad}
\end{figure}

We present the ratio between cooling radii and stellar half mass radii in \fig{cool_rad}. Here we see a number of features of note. Firstly, there is a low mass distribution ($M_\star \lesssim 10^{8.8}$ M$_\odot$) with a large scatter centred on a ratio of unity; these are the diffuse clumpy systems prevalent at low stellar masses that are not undergoing a transition in size. There is then a second distribution of galaxies with stellar masses in the range $10^{8.8} \lesssim \ $M$_\star/$M$_\odot \lesssim 10^{9.8}$ with a negative relation between the ratio of cooling radius and stellar half mass radius; these are galaxies where efficient localised cooling has begun, allowing gas to collapse to high densities below the size of the stellar distribution. Once collapsed to sufficient density these localised cool regions enable high sSFRs, which lead to the significant stellar mass growth associated with decreases in stellar size (\fig{hmr_ssfr}). These galaxies are the galaxies with decreasing gas and stellar sizes evident in \fig{hmr_star_gas_comp}. Once enough stellar mass has been formed in the compact star forming gas distribution the gas and stellar size measurements become comparable and the ratio tends towards unity. This process has been completed for most galaxies by the time they reach stellar masses of $10^{10}$ M$_\odot$. 

These distributions are exemplified by the curve showing the 50$^\mathrm{th}$ percentile. The trend starts at a ratio of unity at $M_\star=10^8$ M$_\odot$ and then exhibits a clear dip in the stellar mass regime dominated by galaxies undergoing the transition from diffuse to compact stellar distributions. The trend then returns to a ratio of unity for galaxies at stellar masses $>10^{10}$ M$_\odot$.

We can also see a number of outliers to this trend with $R_{\mathrm{gas},1/2} / R_{\star,1/2} > 1$ and $M_\star > 10^{9.5} \mathrm{M}_\odot$. These are massive galaxies with compact bulges, where regions of their extended gas distribution have begun collapsing, leading to the transition from compact to extended sources. We discuss this process in detail in \sec{evolve}.

The reason for the specific stellar mass at which we see the diffuse to compact transition is now clear: it is this mass at which gas is capable of cooling efficiently enough to form highly localised regions of sufficient sSFR to form compact stellar systems. We note that, although we cannot show
the effects of resolution on this transition\footnote{The existing EAGLE high resolution simulation (denoted RECAL) is too small in volume (25$^{3}$ cMpc$^{3}$) to form massive compact galaxies at high redshift.}, it takes place significantly above the mass resolution of the simulation and thus should be robust to resolution effects. The compact distributions are, however, near the spatial resolution of the simulation; this could place a lower bound on the sizes that are possible in \flares. The resolution of the underlying dark matter distribution could also lead to spurious collisional heating, leading to kinematic and morphological effects \citep[e.g.][]{Wilkinson_22}. We will investigate the effects of resolution explicitly in \flares-2 with a subset of high resolution simulations using the SWIFT open-source simulation code \citep{SWIFT}.

\section{Compact Galaxy Evolution}
\label{sec:evolve}

In this section, we explore how galaxies evolve from compact systems at high redshift, with a negative size-mass relation, to the extended systems prevalent at the present day, with a positive size-mass relation. 

\subsection{Birth Property Evolution}
\label{sec:birthprops}

\begin{figure*}
	\includegraphics[width=\linewidth]{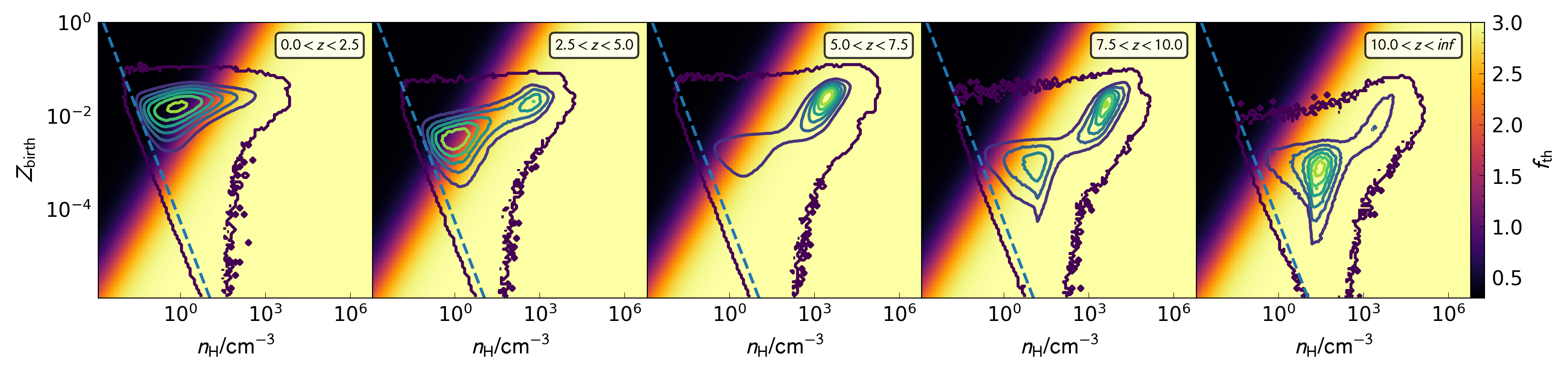}
    \caption{The stellar formation properties that control star formation and feedback in the subgrid model. The contours show the distribution of star formation properties in each redshift bin for all stellar particles split into redshift bins. The background shows the feedback fraction given by \eq{fth}. The dashed line indicates the star formation threshold defined in \eq{star_thresh}. Bins where $z<5$ contain only stellar particles from the \eagle\ AGNdT9 sample while bins where $z>5$ contain both stellar particles from \flares\ and \eagle\ AGNdT9, where the AGNdT9 variant has been used to best match the transition from \flares\ at high redshift to \eagle\ at low redshift.}
    \label{fig:birthmetden}
\end{figure*}

Macroscopic changes in galaxy properties, such as size, are tracers of physical mechanisms taking place at much smaller scales within. Particularly those taking place in their cores at high redshift, as shown in \sec{form}. To probe the small scale physical mechanisms driving galaxy size evolution we present the stellar formation properties of all stellar particles in \flares\footnote{Note that \flares\ is biased towards regions of high overdensity in which the most massive galaxies form at $z>5$. However, these massive galaxies are the subject of this investigation and thus the bias does not affect any conclusions.} 
and \eagle\ AGNdT9 in \fig{birthmetden}. Here we bin these stellar particles by the redshift of their formation to show the redshift evolution of the density and enrichment of star forming environments. The background of \fig{birthmetden} shows the feedback fraction $f_\mathrm{th}$ (described in \sec{eagle_star_fb}) for each combination of birth density and metallicity, while the dashed line shows the star formation threshold.

In the right hand panel of \fig{birthmetden} we can see the first stars to form in the simulation do so at low density and low metallicity ($Z_{\rm birth}\sim10^{-3}$, $n_{\rm H}\sim10^{1}{\rm cm}^{-3}$), with a small contribution by stellar particles forming in the earliest compact cores of massive galaxies at high metallicity and density ($Z_{\rm birth}\sim10^{-1.5}$, $n_{\rm H}\sim10^{4}{\rm cm}^{-3}$). These early cores have already started to cool and collapse at these redshifts, further aiding their enrichment, and creating a feedback loop of star formation and enrichment.

Between $7.5 \leq z < 10$ compact core star formation begins to dominate as the most massive galaxies begin to form their compact cores in earnest. Here, we see a shift from the low metallicity and density locus to one of high density and metallicity corresponding to the star formation taking place in massive galaxies' compact cores. For stellar particles formed between $5 \leq z < 7.5$, when the majority of compact galaxies undergo the transition from diffuse to compact, this high density and metallicity locus dominates with only a small contribution from lower densities. This tail of low density star formation has a tighter distribution in terms of metallicity, a reflection of the enrichment of the wider gas environment from previous star formation. 

From $z=5$ to $z=0$ (two left hand panels of \fig{birthmetden}) the locus of high density and metallicity star formation shifts to low density with a spread in metallicity from intermediate ($Z_{\rm birth}\sim10^{-2.5}$) to high values ($Z_{\rm birth}\sim10^{-2}$). By $0 \leq z < 2.5$ this low density locus is well established, and has shifted to higher values of metallicity ($Z_{\rm birth}\sim10^{-2}$). In this epoch, even gas in low density environments has been significantly enriched enough to enable star formation. 

Recall \fig{gas_hmr_comp}, where the majority of compact galaxies were shown to be enshrouded by extended non-star forming gas distributions. When combined with the evolution of stellar birth properties in \fig{birthmetden} we can see a clear correlation, where low density and low metallicity gas in the extended gas distributions around compact galaxies begins to form stellar particles after $z=5$. This is enabled by the extended gas distributions not only reaching the required densities at a later epoch than the compact cores, but also due to sufficient mixing from the metal enriched core due to increases in AGN and stellar feedback efficiency. This is the onset of inside-out star formation as seen at lower redshift \citep{Tacchella_15a, Tacchella_15b, Nelson_16, Tacchella_18, Nelson_19, Wilman_20, Matharu_22}, but beginning earlier than yet observed at $z=5$.

Focusing on the location of the stellar birth property locus relative to the feedback fraction, $f_\mathrm{th}$, shown in the background of \fig{birthmetden}, we can see that $f_\mathrm{th}$ is maximal for the majority of high redshift star formation. Only at $z<5$ does the locus of star formation include densities and metallicities at which $f_\mathrm{th}$ is not maximal. Despite this, we have shown in \sec{star_fb} that even the maximal feedback fraction in the fiducial \eagle\ model is incapable of affecting the wider surroundings at high redshift. We will investigate the effects of allowing larger feedback fractions in this regime in a future work that modifies the subgrid model.

\subsection{Birth property dependence on large-scale environment}

\begin{figure}
	\includegraphics[width=\linewidth]{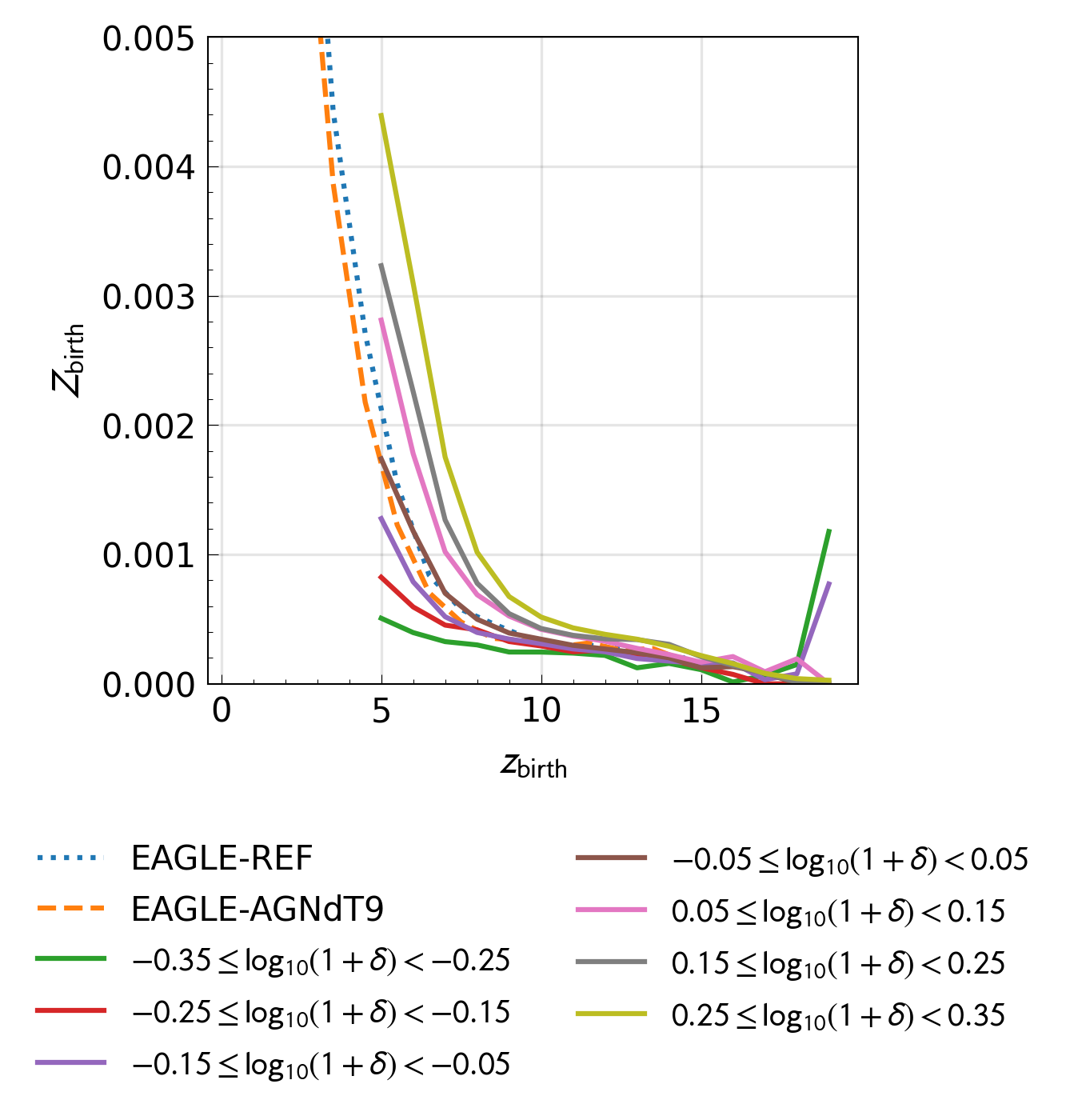}
    \caption{The redshift evolution of stellar birth metallicity in \flares\ (solid lines) and both \eagle\ \texttt{REF} (dotted line) and \texttt{AGNdT9} (dashed line). The \flares\ galaxies are divided into overdensity bins to show the environmental dependence of stellar birth metallicity. Each curve represents the 50$^{\mathrm{th}}$ percentile of the underlying distribution. Unlike the plots of integrated galactic quantities, all stellar particles formed in the simulation are included. We truncate this plot at $Z_\mathrm{birth}=0.005$ to better show the environmental dependence, both \texttt{REF} and \texttt{AGNdT9} continue monotonically increasing to $Z_\mathrm{birth}\sim0.022$ at $z=0$ with \texttt{AGNdT9} forming at marginally higher metallicities than \texttt{REF}.}
    \label{fig:birthmet_evo}
\end{figure}

\begin{figure}
	\includegraphics[width=\linewidth]{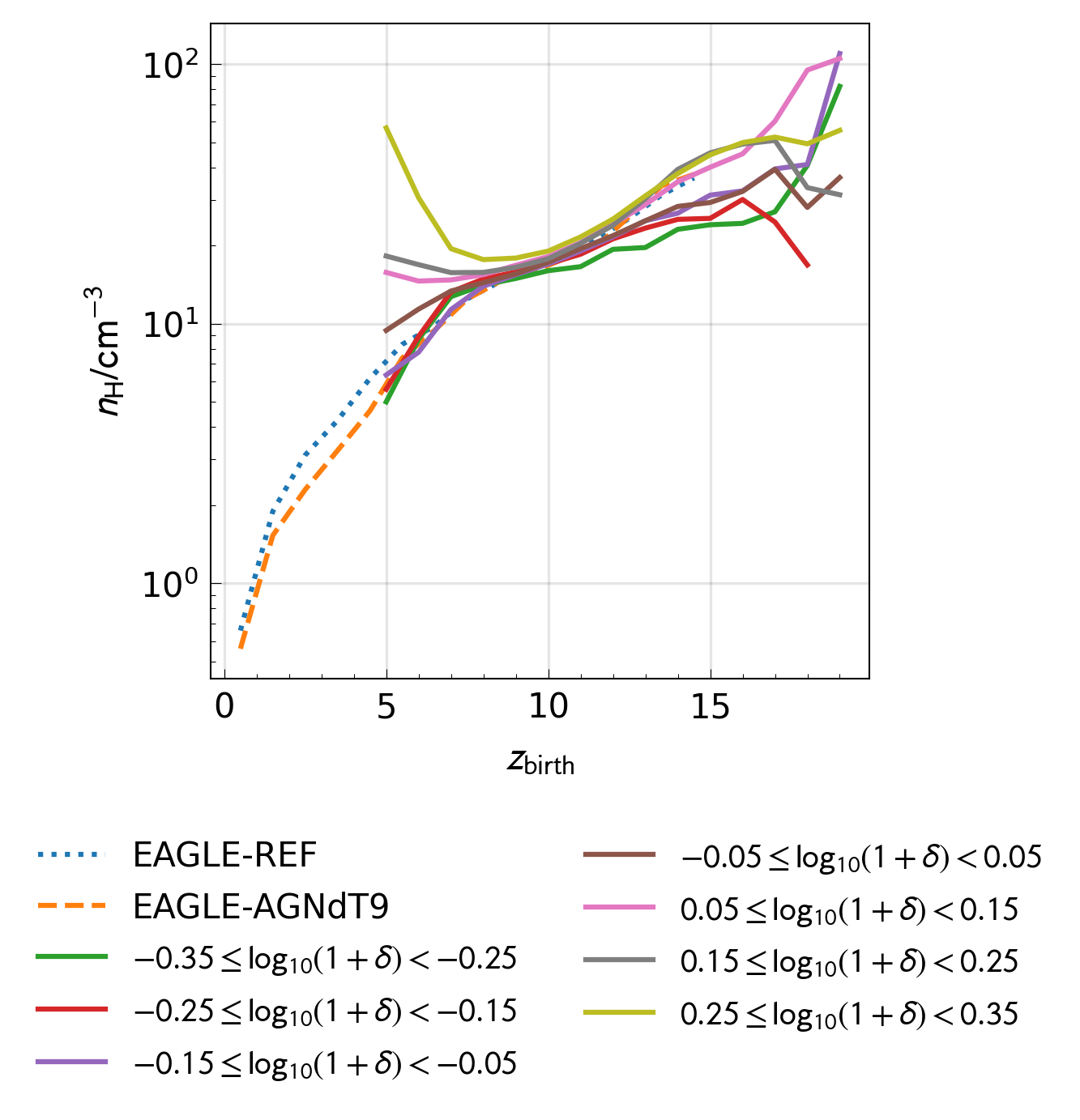}
    \caption{The redshift evolution of stellar birth density in \flares\ (solid lines) and both \eagle\ \texttt{REF} (dotted line) and \texttt{AGNdT9} (dashed line). The \flares\ galaxies are divided into overdensity bins to show the environmental dependence of stellar birth density. Each curve represents the 50$^{\mathrm{th}}$ percentile of the underlying distribution. Unlike the plots of integrated galactic quantities, all stellar particles formed in the simulation are included.}
    \label{fig:birthden_evo}
\end{figure}

As the universe evolves, stellar particles enrich their local environment and stellar birth metallicity increases. One of the main strengths of the \flares\ approach is the high dynamic range of environments covered by the resimulations. Using this we can investigate the birth properties in specific large scale environments, defined by the overdensity of a region (overdensity smoothed over a spherical aperture with a radius of 15 cMpc h$^{-1}$), and shed light on the onset of enrichment and local density in specific environments. In \fig{birthmet_evo} and \fig{birthden_evo} we present the median redshift evolution of stellar birth metallicity and stellar birth density, respectively, for all stellar particles formed by $z=5$ in \flares\ and all stellar particles formed by $z=0$ for both \eagle\ REF and AGNdT9. The \flares\ sample is split into overdensity bins (environments). The \eagle\ samples are representative of mean overdensity environments.

In terms of stellar birth metallicity evolution (\fig{birthmet_evo}) we find little distinguishes each environment at the earliest times ($z\geq17$) beyond stochastic star formation where a small number of stars are formed. Following this, the most over-dense regions with $\log_{10}(1+\delta)> 0.05$ begin forming enriched stellar particles after $z\sim15$. From $z\sim10$ onwards the environmental dependence on stellar birth metallicity is well established, with the enrichment of new stellar particles at fixed redshift increasing with increasing overdensity. This environmental dependence can be interpreted as delayed enrichment of star forming gas in lower overdensity environments relative to more over-dense environments. The most massive and early forming galaxies are strongly biased towards the highest overdensities, hence this dependence on environment. The \eagle\ curves are both consistent with the mean overdensity bins from \flares, yielding a strong evolution in stellar birth metallicity which is mimicked at earlier times by the most over-dense regions.

In contrast to stellar birth metallicity, stellar birth density (\fig{birthden_evo}) exhibits environmental dependence at early times beyond stochasticity. As early as $z=15$ we can see under-dense regions form stars at noticeably lower densities than over-dense environments. This naturally follows from the definition of an environment by overdensity; a more over-dense region contains more mass, and thus achieves higher densities earlier. By $z=10$ this early environmental dependence is no longer present. Early enrichment in the over-dense regions (see \fig{birthmet_evo}) aids low density star formation, allowing lower stellar birth densities. The early onset and loss of the environmental dependence of stellar birth density due to enrichment at $z>10$ shows a clear transition between regimes where gas density, and later gas metallicity, dominate the star formation law. 

We then see a strong environmental dependence establish itself once again in \fig{birthden_evo} between $z\sim8-9$ as the compact cores of the most massive galaxies, biased towards the most over-dense regions, begin to form in earnest. In the lowest density environments, these galaxies do not form by $z=5$ and, even in mean density regions, they are few in number at this redshift\footnote{\flares\ contains infinitely many more $M_{\star} / M_\odot > 10^{10}$ galaxies at $z=5$ than the periodic \eagle\ simulations.}. This explains the strong dependence on environment present at $z=5$, with the most over dense regions increasing in stellar birth density while lower density regions exhibit a decrease. From $z=5$ onwards we can see a clear evolution towards lower stellar birth density as star forming gas is enriched, enabling star formation at lower densities. This is also the regime where the low density extended gas distributions begin to form stars, causing the increase in size of massive compact galaxies.

\subsection{Spatial Distribution Evolution}

\begin{figure}
    \includegraphics[width=\linewidth]{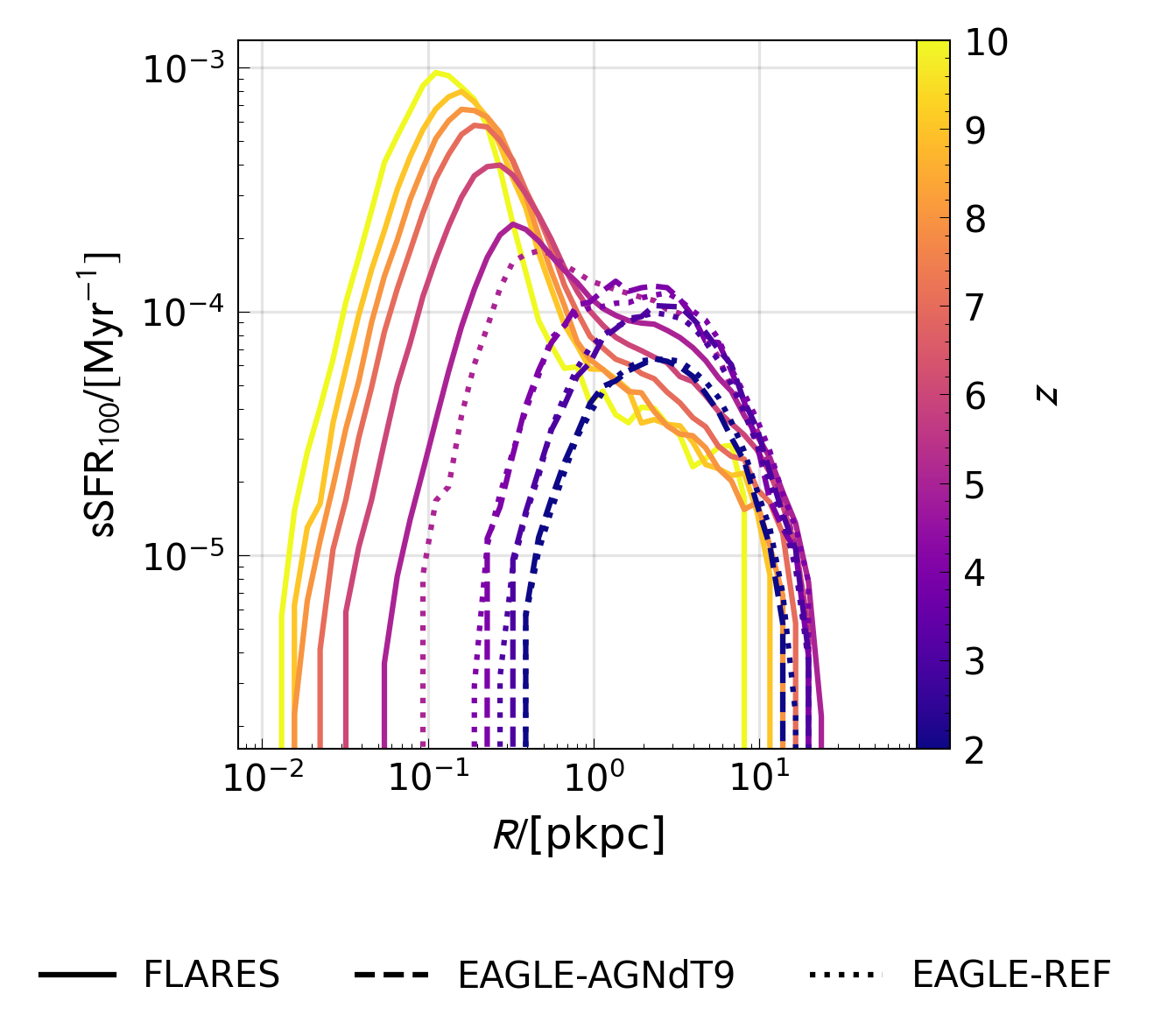}
    \caption{Specific star formation rate (sSFR) profiles binned by redshift for all \flares\ galaxies at $z\geq5$ (solid lines), all \eagle-\texttt{REF} galaxies at $z\leq5$ (dotted lines), and all \eagle-\texttt{AGNdT9} galaxies at $z\leq4$ (dashed lines) with $M_\star/M_\odot\geq 10^{9}$. Each curve represents the 50$^{\mathrm{th}}$ percentile of all sSFR profiles in a redshift bin. \eagle-\texttt{AGNdT9} $z=5$ has too few galaxies in this mass/redshift regime due to its volume and was thus omitted.}
    \label{fig:sfr_prof}
\end{figure}

To explicitly link the shift of star formation to a low density and moderate metallicity locus at $z<5$ to star formation in extended gas distributions around compact galaxies, we investigate the spatial distribution of star formation. To do so we calculate profiles of sSFR (relative to the stellar mass of the whole galaxy) at specific redshifts by binning all stellar particles formed in the last 100 Myrs before a snapshot into annuli. We present the median specific star formation rate (sSFR) profiles for galaxies with $M_\star/M_\odot\geq 10^{9}$ in each snapshot ($0 \leq z \leq 10$) in \fig{sfr_prof}. 

At ($z>5$) there is a clear peak of star formation at small radii, corresponding to compact core star formation in massive galaxies. This peak of star formation decreases with decreasing redshift as the compact cores become less dominant. For $z<6$ there is an increase in the star formation taking place at $R>1$ ckpc. By $z\lesssim4$ the peak of core star formation has entirely disappeared and is replaced by a peak at $R>1$ ckpc. This is a direct representation of the shift from the high redshift regime, where compact core star formation is dominant, to the low redshift regime, where extended gas distribution star formation dominates. This shift between regimes yields the transition shown in \fig{star_hmr} from a negative size-mass relation to a positive size-mass relation. We can summarise these two regimes as a high redshift ($z>5$) epoch of compact core (or bulge) star formation and a low redshift ($z<5$) epoch of inside-out star formation.

\section{Conclusions}
\label{sec:conclusion}

In this work, we have demonstrated the physical mechanisms in \flares\ and the \eagle\ sub-grid model that drive the formation of the compact galaxies which drive the negative slope of the size-mass relation at $z\geq5$. Specifically, we find:

\begin{itemize}
    \item There are two formation paths for massive compact galaxies at $z \geq 5$. The compact formation path for galaxies forming at the earliest epochs ($z>10$) and the diffuse formation path for galaxies forming at later times ($z<10$).
    \item In the compact evolution path, progenitors of massive compact galaxies at $z=5$ form at the earliest times ($z>10$) in pristine environments with little to no metal enrichment. Due to the lack of enrichment they form stars at high densities and thus begin as low mass compact systems that stay compact for their entire evolution to $z=5$. 
    \item  In the diffuse evolution path, progenitors of massive compact galaxies at $z=5$, forming after gas has become partially enriched ($z<10$), are capable of forming stars at lower densities, and thus begin as low mass diffuse systems which become compact at later times. 
    \item The transition from diffuse to compact galaxy is driven by runaway star formation in their cores. Gas in a region of the diffuse galaxy begins efficient cooling, reaching high densities, enabling highly localised efficient star formation. This star formation then enriches the gas in the core of the galaxy, further enhancing star formation. Once this process has taken place, the galaxy has a compact stellar core with a high specific star formation rate and thus a compact stellar distribution. 
\end{itemize}

We have also presented an evolutionary path for compact galaxies forming at $z\geq5$, explaining how these galaxies evolve to yield extended sources seen at the present day, and the shift from the surprising negative high redshift stellar size-mass relation to the observed positive relation at lower redshifts. We find:  

\begin{itemize}
    \item Galaxies at $z>5$ with compact stellar distributions are, in the majority of cases, surrounded by non-star forming gas distributions up to $\sim100$ times larger than the stellar component. These extended gas distributions become enriched at later times and become capable of forming stars, increasing the stellar size of galaxies. 
    \item There are two broad regimes of star formation: a high redshift ($z>5$) regime dominated by compact core star formation and a low redshift ($z<5$) regime of inside-out star formation.
    \item Star formation in the compact core regime takes place in gas at high density and high metallicity between $5 \leq z \leq 10$. 
    \item Star formation in the inside-out regime takes place in gas at low density and moderate to high metallicity between $0 \leq z \leq 5$.
    \item A minority of $z>5$ compact stellar distributions are associated with comparably  compact gas distributions. These galaxies are neither excessively young nor numerical artefacts of the halo finding process. The vast majority are satellite systems however a small number are central galaxies. These compact galaxies could be the progenitors of so called `red nuggets' observed at intermediate redshifts.
\end{itemize}

We probe the environmental dependence of stellar birth properties and find a strong dependence rooted in the environmental dependence of galaxy stellar mass. At fixed redshift, high overdensity regions form stellar particles at higher densities (starting at $z\sim8-9$) and higher metallicities (starting at $z\sim10$) than lower overdensity regions. The most overdense regions also undergo a period of high density star formation at early times ($12 \lesssim z \lesssim 20$). At $5 \leq z \lesssim 7$ the highest overdensity regions strongly deviate from the mean and low overdensity regions, forming stars at increasing density while all other regions begin to form stars at decreasing density.

Given the wealth of observational data coming from JWST, this insight into the physical mechanisms governing the formation and evolution of galaxies in the Epoch of Reionisation is invaluable to interpreting upcoming observational samples, enabling the mapping from observational properties to underlying physical mechanisms. Combined with the predicted size-luminosity relations across the spectrum in \cite{Roper22}, the mechanisms detailed in this work can be used to shed light on the formation and evolution of galaxies at the highest redshifts, linking the low redshift Universe to the high redshift Universe.

Not only are these insights important to interpreting observations, they are also invaluable to future sub-grid models, particularly if the negative size-mass relation is found to be isolated to simulations. Current models are designed to work well at low redshift \citep[e.g.][]{schaye_eagle_2015, crain_eagle_2015, Furlong_2017} and, as shown in previous \flares\ studies amongst many others, perform surprisingly well at high redshift. Regardless of this favourable performance, given the ever increasing frontier of high redshift astronomy, having robust high redshift motivated models for galaxy formation in simulations is imperative to keep up with the observed samples, and to aid their interpretation. Understanding the behaviour of modern sub-grid models in this epoch is an important first step towards achieving this. 

\section*{Acknowledgements}

We thank the \eagle\ team for their efforts in developing the \eagle\ simulation code.
We also wish to acknowledge the following open source software packages used in the analysis: \textsf{scipy} \citep{2020SciPy-NMeth}, \textsf{Astropy} \citep{robitaille_astropy:_2013}, \textsf{CMasher} \citep{Cmasher}, and \textsf{matplotlib} \citep{Hunter:2007}.

This work used the DiRAC@Durham facility managed by the Institute for Computational Cosmology on behalf of the STFC DiRAC HPC Facility (www.dirac.ac.uk).
The equipment was funded by BEIS capital funding via STFC capital grants ST/K00042X/1, ST/P002293/1, ST/R002371/1 and ST/S002502/1, Durham University and STFC operations grant ST/R000832/1.
DiRAC is part of the National e-Infrastructure. The \eagle\ simulations were performed using the DiRAC-2 facility at Durham, managed by the ICC, and the PRACE facility Curie based in France at TGCC, CEA, Bruyeres-le-Chatel.

WJR acknowledges support from an STFC studentship. 
CCL acknowledges support from a Dennis Sciama fellowship funded by the University of Portsmouth for the Institute of Cosmology and Gravitation.
DI acknowledges support by the European Research Council via ERC Consolidator Grant KETJU (no. 818930).
The Cosmic Dawn Center (DAWN) is funded by the Danish National Research Foundation under grant No. 140.

We list here the roles and contributions of the authors according to the Contributor Roles Taxonomy (CRediT)\footnote{\url{https://credit.niso.org/}}.
\textbf{William J. Roper}: Conceptualization, Data curation, Methodology, Investigation, Formal Analysis, Visualization, Writing - original draft.
\textbf{Christopher C. Lovell, Aswin P. Vijayan}: Data curation, Writing - review \& editing.
\textbf{Dimitrios Irodotou, Jussi Kuusisto, Jasleen Matharu, Louise Seeyave}: Writing - review \& editing.
\textbf{Peter Thomas}: Conceptualization, Resources, Writing - review \& editing.
\textbf{Stephen M. Wilkins}: Conceptualization, Writing - review \& editing.

\section*{Data Availability}
A portion of the data used to produce this work can be found online: \href{https://flaresimulations.github.io/data.html}{flaresimulations.github.io/data}. Much of the analysis used the raw data produced by the simulation which can be made available upon request.
All of the codes used for the data analysis are public and available at \href{https://github.com/WillJRoper/flares-sizes-phys}{github.com/WillJRoper/flares-sizes-phys}.



\bibliographystyle{mnras}
\bibliography{flaresIX} 




\appendix

\section{The Effects of AGN on Compact Galaxy Formation}

\begin{figure*}
	\includegraphics[width=\linewidth]{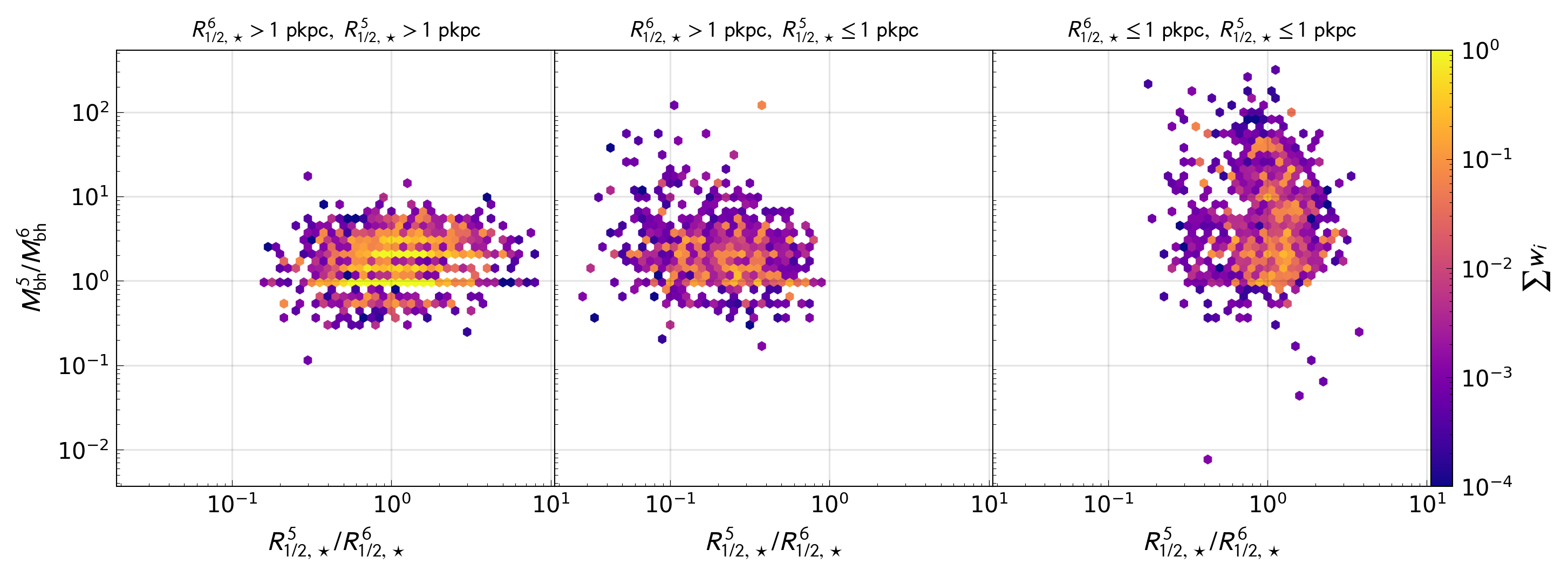}
    \caption{Change in stellar half mass radii vs change in black hole mass between galaxies at $z=5$ and their main progenitors at $z=6$. The change in black hole mass is defined as the sum of all black hole mass present in a galaxy to highlight mass increase driven by accretion over increases due to merger events. Each panel shows galaxies undergoing different phases of size evolution, left to right: galaxies which were diffuse at $z=6$ and remain diffuse at $z=5$, galaxies which were diffuse at $z=6$ and are compact at $z=5$, and galaxies which were compact at $z=6$ and remain compact at $z=5$. Hexbins are coloured by weighted number density using the \flares\ region weighting scheme.}
    \label{fig:bh_deltasize}
\end{figure*}

In this section, we present the effects of AGN feedback on the evolution of galaxies from diffuse systems to compact systems. To do so we need a measure of the integrated energy injected by the AGN. In our model, AGN feedback is closely tied to the accretion rate of a black hole, with the energy injection rate from black hole accretion given by
\begin{equation}
    \dot{E} = \epsilon_\mathrm{r} \epsilon_\mathrm{f} \dot{m}_\mathrm{accr} c^{2},
\end{equation}
\noindent
where $\epsilon_\mathrm{f} = 0.15$ is the fraction of the feedback energy that is coupled with the ISM, $\epsilon_\mathrm{r}=0.1$ the radiative efficiency of the accretion disc, $\dot{m}_\mathrm{accr}$ the (instantaneous) black hole accretion rate and $c$ the speed of light.
We can probe the effect of AGN feedback by measuring the mass increase between snapshots of a black hole and comparing that to the ratio between sizes. However, purely looking at the mass increase of the central black hole will contaminate the mass increase with black hole mergers. To avoid this we look at the ratio between the total black hole mass of a galaxy and its progenitor as inferred from merger graphs \citep{Roper_20}. Note that, despite this, the black hole mass ratios will nonetheless be contaminated by galaxy mergers, however, the vast majority of these contribute seed mass black holes which will have little effect on the mass increase. Galaxy mergers will also produce different stellar size ratios than stellar formation driven changes and thus should not affect the overall trend. 

We present the comparison between stellar size ratios and black hole mass ratios in \fig{bh_deltasize}. Here we see no trend between the stellar size ratio and black hole mass ratio in any size evolution phase (any column). We can thus conclude that AGN and their feedback do not play a role in the compactification of galaxies in the \eagle\ model. The only trend we do see is that a subset of galaxies which were compact at $z=6$ and remain compact $z=5$ exhibit large increases in black hole mass. These are galaxies with black holes efficiently accreting from the dense gas in their compact cores, and are thus examples of possibly observable high redshift quasars.


\bsp	
\label{lastpage}
\end{document}